\documentclass[12pt, a4paper]{article}
 \usepackage[font=small,format=plain,labelfont=bf,up,textfont=normal,up,justification=justified,singlelinecheck=false]{caption}

\usepackage[a4paper, left=2cm,right=2cm,top=3.5cm, bottom=3.5cm]{geometry}
\usepackage[colorlinks=true,linkcolor=black,citecolor=teal,urlcolor=MidnightBlue,filecolor=black]{hyperref}
\usepackage{amsfonts}
\usepackage{amsmath}
\usepackage{setspace}
\usepackage[dvipsnames]{xcolor}
\definecolor{SchoolColor}{rgb}{0.6471, 0.1098, 0.1882} 

\usepackage[utf8,applemac]{inputenc}
\usepackage{tensor}
\usepackage{cite}
\usepackage{tikz}
\usepackage{graphicx}
\usepackage{bm} 

\usepackage{dcolumn}
\usepackage{bm}

\newcommand{\bea}{\begin{eqnarray}}
\newcommand{\eea}{\end{eqnarray}}
\newcommand{\be}{\begin{equation}}
\newcommand{\ee}{\end{equation}}
\def\nn{\nonumber}

\def\p{\partial}
\def\eps{\epsilon}

\newcommand{\beqs}{\begin{eqnarray}}
\newcommand{\eeqs}{\end{eqnarray}}
\DeclareMathOperator{\extdm}{d}
\newcommand{\extd}{\extdm \!}

\begin{document}
\begin{titlepage}

\begin{flushright}\vspace{-3cm}
{\small
\today }\end{flushright}
\vspace{0.5cm}

\begin{center}
{ \LARGE{\bf{Invariance of Unruh and Hawking radiation \\ \vspace{10pt} under matter-induced supertranslations}}}
 \vspace{5mm}

\vspace{2mm} 
\centerline{\large{\bf{Geoffrey Comp\`{e}re${}^\dagger$\footnote{e-mail: gcompere@ulb.ac.be}, Jiang Long${}^\star$\footnote{e-mail: jiang.long@apctp.org}, Max Riegler${}^{\dagger,\ddagger}$\footnote{e-mail: max.riegler@ulb.ac.be}}}}

\vspace{2mm}
\normalsize
\bigskip\medskip
\textit{${}^\dagger$ Universit\'{e} Libre de Bruxelles and International Solvay Institutes\\
CP 231, B-1050 Brussels, Belgium\\
\vspace{2mm}
}
\textit{${}^\star$ Asia Pacific Center for Theoretical Physics\\
Pohang 37673, Korea\\
\vspace{2mm}
}
\textit{${}^\ddagger$ Center for Gravitational Physics, Yukawa Institute for Theoretical Physics (YITP)\\Kyoto University, Kitashirakawa Oiwakecho, Sakyo-ku, Kyoto 606-8502, Japan\\
\vspace{2mm}
}
\vspace{25mm}

\begin{abstract}
\noindent
Matter fields are supertranslated upon crossing a shock wave, which leads to entanglement of the quantum vacuum between the two regions on either side of the shock wave. We probe this entanglement for a scalar field in a planar shock wave background by computing the Bogoliubov transformation between the inertial and uniformly accelerated observer. The resulting Bogoliubov coefficients are shown to reproduce the standard Unruh effect without dependence on the form factor of the shock wave. In contrast, excited states lead to observables that depend upon the form factor. In the context of nonspherical gravitational collapse, we comment that the angular dependence of the limiting advanced time leads to similar supertranslation effects that do not affect the Hawking spectrum but do affect scattering amplitudes.

\end{abstract}


\end{center}

\end{titlepage}

\newpage
\tableofcontents
\hypersetup{linkcolor=SchoolColor}
\newpage

\section{Introduction}

While a quantum completion of Einstein gravity in four-dimensional asymptotically flat spacetimes that would allow detailed computations is still far from reach, describing universal properties of Einstein gravity as a quantum low energy effective theory is possible with current techniques. The laws of black hole thermodynamics are early milestones in this endeavor, which, in particular, include Hawking's radiation and paradox \cite{Hawking:1974sw} and the Bekenstein-Hawking area law \cite{Bekenstein:1972tm,Bekenstein:1973ur,Hawking:1974sw}. Another universal result is the graviton dominance of high energy scattering at small deflection angles \cite{tHooft:1987vrq,Amati:1987uf,Banks:1999gd,Amati:2007ak,Ciafaloni:2015xsr} that is described classically by a collision of shock waves \cite{Aichelburg:1970dh}. Recently, a new universal property of gravity was uncovered in the infrared sector close to null infinity \cite{Strominger:2013jfa,He:2014laa,Strominger:2014pwa}: a triangular relationship exists between the displacement memory effect \cite{Zeldovich:1974aa,Christodoulou:1991cr,Bieri:2010tq,Bieri:2013ada}, the leading soft graviton theorem \cite{Weinberg:1965nx} and supertranslation asymptotic symmetries \cite{Bondi:1962px,Sachs:1962zza,Sachs:1962wk}, see the reviews \cite{Strominger:2017zoo,Compere:2018aar}. 

Displacement memory is caused by hard (i.e. finite energy) processes reaching null infinity that can be either a change of Bondi mass \cite{Zeldovich:1974aa}, null matter radiation \cite{Bieri:2010tq,Bieri:2013ada} or gravitational waves \cite{Christodoulou:1991cr}. Displacement memory effectively amounts to a shift of a canonical variable defined at null infinity by a supertranslation asymptotic symmetry \cite{Strominger:2014pwa}. As a result, hard processes at null infinity are accompanied by soft (i.e. infinitely low energy) processes that are mimicked by a zero energy process: a supertranslation asymptotic symmetry. This mimicking can arise because of the infrared limit taken at null infinity. In the bulk of spacetime, a supertranslation can be induced by a shock wave \cite{Dray:1984ha}. For planar shock waves, probes encounter a permanent displacement shift upon crossing. This has led to \emph{define} the concept of soft radiation away from the infrared limit as (exactly zero energy) large diffeomorphisms (or large gauge transformations) acting in the bulk of spacetime. In particular, it led to the definition of black holes with soft hair implants \cite{Donnay:2015abr,Hawking:2016msc,Compere:2016jwb,Compere:2016hzt,Averin:2016ybl,Hawking:2016sgy,Mao:2016pwq,Averin:2016hhm,Donnay:2016ejv,Kolekar:2017yoi,Zhang:2017geq,Mishra:2017zan,Gomez:2017ioy,Chatterjee:2017zeb,Kolekar:2017tge,Chu:2018tzu,Kirklin:2018wvq,Cvetkovic:2018dmq,Averin:2018owq,Choi:2018oel,Donnay:2018ckb,Haco:2018ske,Haco:2019ggi}. 

Furthermore, it has been pointed out that soft hair may bear on the black hole information paradox \cite{Strominger:2014pwa,Hawking:2016msc} and that quantum purity might be restored in principle by correlations between the hard and soft radiation \cite{Strominger:2017aeh} (see also comments in \cite{Donnelly:2017jcd,Marolf:2017jkr,Compere:2019ssx}). While it has been shown that the soft particles factor out of the S-matrix \cite{Mirbabayi:2016axw,Bousso:2017dny,Bousso:2017rsx} (see also \cite{Gabai:2016kuf}), it was also pointed out that outgoing hard and soft states are highly correlated \cite{Carney:2017jut}. A resolution of the information paradox with soft hair would require a conjectured one-to-one correlation between hard and soft states while not affecting the late time Hawking spectrum \cite{Strominger:2017aeh}. During the course of this work, the paper \cite{Javadinazhed:2018mle} appeared where the authors showed that the spectrum of Hawking radiation (without backreaction) emitted in the Schwarzschild background is unchanged after including the dressing of asymptotic states with soft form factors. 

Complementary to this analysis, we derive in this paper the effects of soft hair implants by addressing the question: ``\emph{How do supertranslations caused by hard processes affect Unruh and Hawking radiation?}''. This complements most of the literature on soft factors by focusing on bulk matter-induced supertranslation effects instead of asymptotic soft effects (for closely related work, see \cite{Kolekar:2017yoi,Kolekar:2017tge}).

This paper is organized as follows. In Sec.~\ref{sec:quan} we review the quantization of  a massless complex scalar field in the presence of a planar gravitational shock wave following \cite{Klimcik:1988az} (see also   \cite{Gibbons1975,Hortacsu:1990ku,Hortacsu:1992vu,Dorca:1993sv,Dorca:1994pf,Dorca:1996yt,Hortacsu:1998sw}). We provide two distinct definitions of a complete set of globally well defined planar waves and derive their orthogonality relations under the Klein-Gordon inner product. In Sec.~\ref{sec:Rindler} we make the Rindler observers moving at uniform acceleration explicit and define their associated Rindler modes. In Sec.~\ref{sec:Bogoliubov}, we obtain the Bogoliubov coefficients that map the global shock wave vacua to the left and right Rindler vacua. We compute the resulting spectrum and show that it agrees with the standard Unruh result. We also analyze the occupation numbers for excited states. In Sec. \ref{sec:BH}, we review Hawking's particle production and discuss the parallels between the limiting advanced time shift and a supertranslation. In addition, we comment on the dependence of scattering on the factor characterizing nonspherical collapse. Our conclusion can be found in Sec.~\ref{sec:Discussion}. We collected the proof of the orthogonality of planar waves in the shock wave background in Appendix~\ref{sec:ortho}.

\section{Global plane waves in a shock wave background}
\label{sec:quan}

Following Dray and 't Hooft \cite{Dray:1984ha} we consider a $D$-dimensional planar shock wave in Minkowski spacetime
    \begin{subequations}\label{ds2}
        \begin{align}
        \extd s^2 &= -\extd u \extd v + f(\vec{x}) \delta(u-u_0) \extd u^2+\sum_{i=2}^{D-1} (\extd x^i) ^2, \\
        &= -\extd u \extd\hat v - \Theta(u-u_0) \p_i f(\vec{x}) \extd u\extd x^i +\sum_{i=2}^{D-1} (\extd x^i) ^2 .  
        \end{align}
    \end{subequations}
Here we used the light-cone coordinates $u=t-z$, $v=t+z$ and the $D-2$ transverse coordinates $\vec{x}\equiv (x^2,\cdots,x^{D-1})$. The function $f(\vec{x})$ is the shock wave form factor. The alternative coordinates ${u}=\hat t-\hat z$, $\hat v=\hat t+\hat z$ are obtained by the ``planar supertranslation'' shift
    \begin{equation}
        \hat v = v - \Theta(u-u_0) f(\vec{x}).
    \end{equation}
The stress-tensor has only one nonzero component, $T_{uu}=\delta(u-u_0) T(\vec{x})$, localized on the shock wave front $u=u_0$ that moves towards the positive $z$ direction. Einstein's equations for this spacetime reduce to 
\mbox{$\Delta f(\vec{x}) = -16\pi G_N\,T(\vec{x})$} where $\Delta$ is the $(D-2)$-dimensional Laplacian. The solution is $D$-dimensional Minkowski spacetime outside of $u = u_0$. Causality requires that the Shapiro delay \cite{Shapiro:1964uw} of a probe particle is non-negative and therefore requires $f(\vec{x})$ to be a non-negative function \cite{Camanho:2014apa}. 


Note that the metric \eqref{ds2} is exactly
    \begin{equation}
        g_{\mu\nu}=\eta_{\mu\nu} + \Theta(u-u_0) \mathcal L_{\xi_T} \eta_{\mu\nu},
    \end{equation}
where $\xi_T = f(\vec{x})\p_{\hat v}$, which is similar to a Schwarzschild black hole with a linearized shock wave, see (5.14) of \cite{Hawking:2016sgy}. Here, the metric $g_{\mu\nu}$ is a complete non-linear solution to Einstein's equations.

\subsection{Global plane wave solutions}\label{sec:GlobalPlaneWave}

The quantization of a massless Klein-Gordon field $\Phi$ with $D$-momentum $k_\mu$ and components $k_t,\,k_{z},\,\vec{k}$ in this background is reviewed in great detail in \cite{Klimcik:1988az}. After decomposing the scalar field in Fourier modes along $v$,  \mbox{$\Phi = e^{-i k_- v}\psi (u,\vec{x})$}, where $k_\pm=\frac{1}{2}(k_t\pm k_{z})$ the Klein-Gordon equation reduces to a Schr\"odinger equation
    \begin{equation}\label{eq:Schrodinger}
        i \p_u \psi =  \left( - \frac{\Delta}{4 k_-} - f(\vec{x}) k_- \delta(u-u_0) \right)  \psi.
    \end{equation}
In the region $u>u_0$ or $u<u_0$ a complete set of modes is given by standard plane waves. The presence of the shock wave introduces a nontrivial junction condition between these plane waves. In fact, the shock wave implies that the modes at $u \rightarrow u_0^-$ are related to the modes at $u \rightarrow u_0^+$ by a ``planar supertranslation'' of the form 
    \begin{equation}
        v \rightarrow v - f(\vec{x}). 
    \end{equation}
Let us choose a momentum eigenstate $f^I_{k_-,\vec{k}}$ in the region $u<u_0$, 
    \begin{equation}\label{Phiun}
        f^I_{k_-,\vec{k}} |_{u<u_0} = e^{-i k_- v}\psi_{u<u_0},\qquad   \psi_{u<u_0}(u,\vec{x} ; k_-,\vec{k}) = N_k e^{-i k_+ (u-u_0) +i \vec{k}\cdot\vec{x}},
    \end{equation}
where $k_+=\frac{\vec{k}^2}{4k_-}$. We call $f^I_{k_-,\vec{k}}$ an ``initial eigenstate'' because for $u\rightarrow-\infty$ it is a planar wave with a unique momentum. As detailed in \cite{Klimcik:1988az}, the field is then given at $u=u_0^+$ by
    \begin{equation}
        f^I_{k_-,\vec{k}} |_{u=u_0^+} = N_k e^{-i k_- v} e^{
        i (\vec{k}\cdot\vec{x} + k_- f(\vec{x}))}. 
    \end{equation}
The field is then propagated for $u>u_0$ using the free wave equation. It is straightforward to show that the global initial-eigenstate solution can be written in the form 
    \begin{equation}\label{solfL}
    f^I_{k_-,\vec{k}}(u,v,\vec{x}) = N_k e^{-i k_- v+ i \vec{k}\cdot\vec{x}}\int \frac{\extd\bm{x}^\prime}{(2\pi)^{D-2}} \int \extd\bm{k}' e^{ i(\vec{k} -\vec{k}^{\prime} )(\vec{x}^{\prime} -\vec{x})-i \frac{\vec{k}^{\prime 2}}{4 k_-} (u-u_0) +i k_- \Theta(u-u_0) f(\vec{x}')}, 
    \end{equation}
where $\extd\bm{x}$ is an abbreviation for the $(D-2)$-dimensional volume differential $\extd x_2\ldots\extd x_{D-1}$ and similarly for $\extd\bm{k}$.

For $u<u_0$, one performs $\int\extd\bm{x}'$ first, which gives $(2\pi)^{D-2}\delta(\vec{k}-\vec{k}')$ and we recover \eqref{Phiun}. For $u>u_0$, one could perform $\int\extd\bm{k}'$ by quadratures. We can also write the solution as
    \begin{subequations}
        \begin{align}
            f^I_{k_-,\vec{k}}(u,v,\vec{x}) &= N_k e^{-i k_- v}  \int \extd\bm{k}' e^{ i \vec{k}^{\prime}\cdot\vec{x} -i \frac{\vec{k}^{\prime 2}}{4 k_-} (u-u_0)} A_{k_-,\vec{k}}(\vec{k}', \Theta(u-u_0)), \label{solG}\\
            A_{k_-,\vec{k}}(\vec{k}', \Theta(u-u_0)) &= \int \frac{\extd\bm{x}^\prime}{(2\pi)^{D-2}} e^{ i(\vec{k} -\vec{k}^{\prime} )\vec{x}^{\prime} +i k_- \Theta(u-u_0) f(\vec{x}') }
= \left\{   \begin{array}{l}  
            \delta(\vec{k}-\vec{k}')\quad \qquad u < u_0, \\
            \phi_{k_-}(\vec{k}-\vec{k}')  \qquad u > u_0,
            \end{array}\right.
        \end{align}
    \end{subequations}
where the mixing factor $\phi_{k_-}(\vec{k}-\vec{k}') $ is defined as
\bea
\phi_{k_-}(\vec{k}-\vec{k}') \equiv \int \frac{\extd\bm{x}^\prime}{(2\pi)^{D-2}} e^{ i(\vec{k} -\vec{k}^{\prime} )\vec{x}^{\prime} +i k_-  f(\vec{x}') }. \label{eq:PhiDef}
\eea

\begin{figure}[t]
		\centering
		\includegraphics[width=\textwidth]{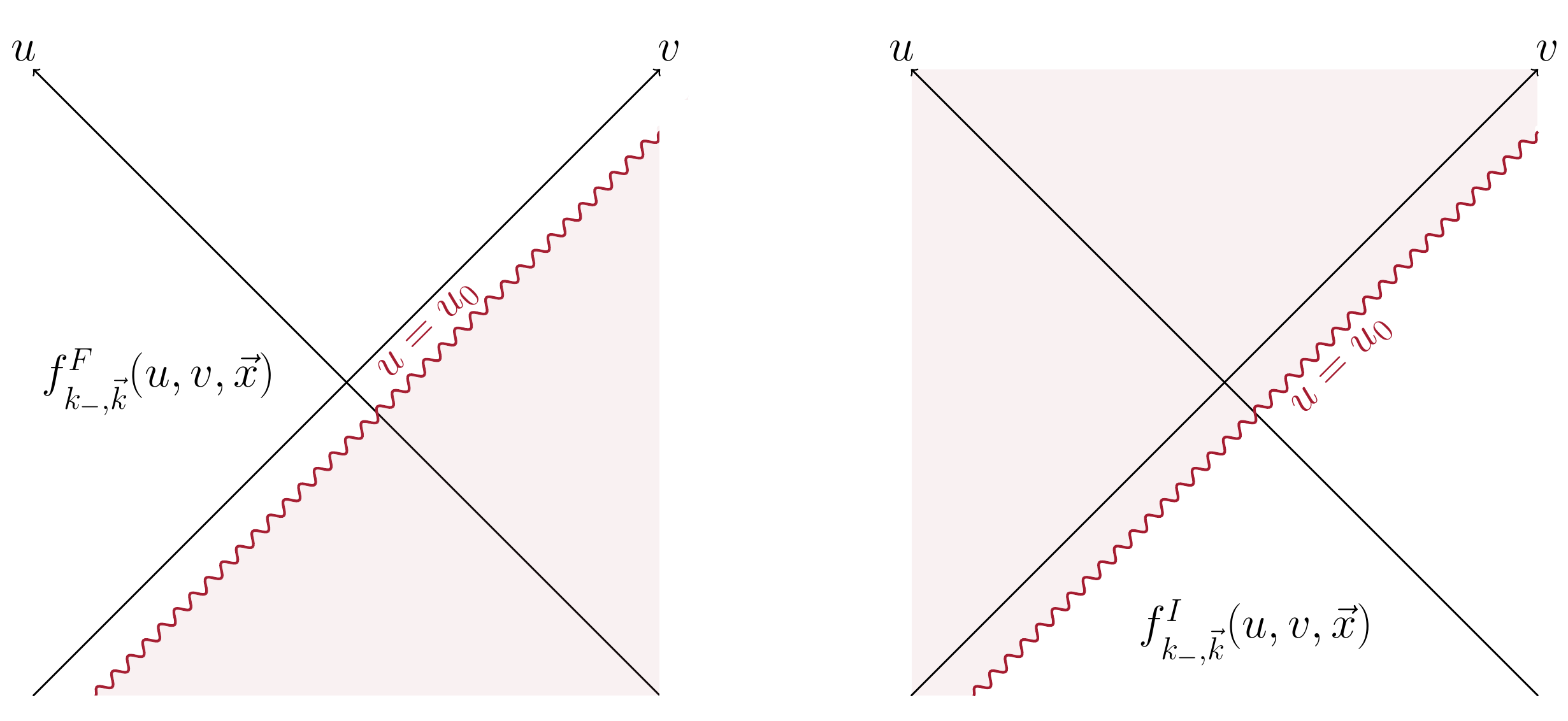}
		\caption{Schematic depiction of initial and final global plane wave eigenstates. The blank (shaded) part depicts the region where each solution is a plane wave eigenstate (a superposition of plane waves).}
		\label{fig:InitialFinalEigenstates}
	\end{figure}

In the case of the Aichelburg-Sexl wave, the integral is elementary, see (10) of  \cite{tHooft:1987vrq}. For future use we note the useful relations
    \begin{subequations}
        \begin{align}
\int \extd \bm{k}'   \phi^*_{k_-}(\vec{k}'-\vec{k})   \phi_{k_-}(\vec{k}'-\vec{l}) &= \delta(\vec{k}-\vec{l}), \\ 
\int \extd\bm{k}'   \phi_{k_-}(\vec{k}'+\vec{k})   \phi_{k_-}(-\vec{k}'-\vec{l}) &= \phi_{2 k_-}(\vec{k}-\vec{l}). 
        \end{align}\label{propphi}
    \end{subequations}
Alternatively, one could have chosen a momentum eigenstate at late retarded times $f^F_{k_-,\vec{k}}(u,v,\vec{x})$ that we will denote as a ``final eigenstate''. Formally, this solution can be obtained from the substitution
    \begin{equation}
        u\rightarrow-u,\qquad v\rightarrow-v,\qquad k_-\rightarrow-k_-,\qquad k_+\rightarrow -k_+,\qquad u_0\rightarrow-u_0,
    \end{equation}
that leaves the metric \eqref{ds2} invariant. Explicitly, we have
    \begin{equation}\label{eq:GlobalMinkEigenFinal}
        f^F_{k_-,\vec{k}}(u,v,\vec{x}) = N_k e^{-i k_- v}  \int \extd\bm{k}' e^{ i \vec{k}^{\prime}\cdot\vec{x} -i \frac{\vec{k}^{\prime 2}}{4 k_-} (u-u_0)} A_{-k_-,\vec{k}}(\vec{k}', \Theta(u_0-u)).
    \end{equation}
See Fig.~\ref{fig:InitialFinalEigenstates} for a schematic depiction of initial and final global plane wave eigenstates. 
Both $f^I_{k_-,\vec{k}}$ and $f^F_{k_-,\vec{k}}$ solve the Schr{\"o}dinger equation \eqref{eq:Schrodinger} and thus provide two schemes to define the shock wave vacuum. For definiteness, we will focus on the initial eigenstate basis $f^I_{k_-,\vec{k}}$ in the following but we will also comment on the analogue results using the final eigenstate basis $f^F_{k_-,\vec{k}}$.  

\subsection{Complete orthonormal basis}

The set of initial eigenstates $f^I_{k_-,\vec{k}}$ written in \eqref{solfL} forms a complete basis of solutions to the Klein-Gordon equation\footnote{We ignore here the measure zero set of plane waves localized at $u=u_0$.}. Therefore, we can decompose the scalar field in terms of these modes as
    \begin{equation}\label{eq:MinkPhi}
        \Phi(u,v,\vec{x})= \sum_{k_- > 0, \vec{k}} \left( a_{k_-,\vec{k}} f^I_{k_-,\vec{k}} (u,v,\vec{x})+ a^\dagger_{k_-,\vec{k}} f^{*I}_{k_-,\vec{k}}(u,v,\vec{x}) \right). 
    \end{equation}
After promoting the coefficients $a_{k_-,\vec{k}}$ and $a^\dagger_{k_-,\vec{k}}$ to operators and interpreting them as annihilation and creation operators, respectively, one can define the initial global shock wave vacuum $|0_{\textrm{SW}}\rangle_I $ as
    \begin{equation}
        a_{k_-,\vec{k}}|0_{\textrm{SW}}\rangle_I = 0,
    \end{equation}
with $k_- > 0$ and $\vec{k}$ arbitrary.

In addition to forming a complete set of solutions one can, in fact, also show that the modes $f^I_{k_-,\vec{k}}$ form an orthonormal set under the Klein-Gordon inner product that is given by 
    \begin{equation}\label{eq:KGNorm}
        (\Phi_1,\Phi_2)_{KG} = i \int_\Sigma \extd z \extd\bm{x} \sqrt{-g} n_\mu \left(\Phi_1^* \p^\mu \Phi_2 - \Phi_2 \p^\mu \Phi_1^* \right),
    \end{equation}
where $\Sigma$ is a constant Cauchy surface and $n^\mu$ its unit normal vector. If we choose for example the Cauchy slices along $n_\mu = \p_\mu t$ we get
    \begin{equation}
        (\Phi_1,\Phi_2)_{KG} = i \int_\Sigma \extd z \extd\bm{x} \Big( \Phi_1^* \left( \p_u \Phi_2 + (1+2 f(x^i) \delta(u-u_0) ) \p_v \Phi_2 \right)- (\textrm{other\, term}) \Big).
    \end{equation}
It is straightforward but lengthy to show that using this inner product the initial eigenstate modes satisfy
    \begin{subequations}
        \begin{align}
        (f^{*I}_{k_-,\vec{k}}, f^I_{l_-,\vec{l}})_{KG} &=  0,\label{rel1} \\
        (f^{I}_{k_-,\vec{k}}, f^I_{l_-,\vec{l}})_{KG} &= \delta(\vec{l}-\vec{k})\delta(l_z-k_z),\label{rel2} \\ 
        (f^{*I}_{k_-,\vec{k}}, f^{*I}_{l_-,\vec{l}})_{KG} &= -\delta(\vec{l}-\vec{k})\delta(l_z-k_z).\label{rel3}
        \end{align}
    \end{subequations}
Indeed, a positive null frequency plane wave mode $k_- >0$ implies that the plane wave has positive time frequency $k_t = k_+ + k_- >0$. Since null frequencies are untouched by the shock wave, positive frequency modes and negative frequency modes are not mixed by the shock wave, which implies \eqref{rel1} \cite{tHooft:1987vrq,Klimcik:1988az}. The other two orthogonality conditions \eqref{rel2}-\eqref{rel3} require explicit computations that are relegated to Appendix \ref{sec:ortho}. This is a new result. 

The same arguments can be repeated for the final eigenstates $f^F_{k_-,\vec{k}}$. That is, one can decompose the scalar field in terms of these modes as
    \begin{equation}\label{eq:MinkPhiFinal}
        \Phi(u,v,\vec{x})= \sum_{k_- > 0, \vec{k}} \left( \tilde{a}_{k_-,\vec{k}} f^F_{k_-,\vec{k}} (u,v,\vec{x})+ \tilde{a}^\dagger_{k_-,\vec{k}} f^{*F}_{k_-,\vec{k}}(u,v,\vec{x}) \right).
    \end{equation}
After quantization one can define the global final shock wave vacuum as
    \begin{equation}
        \tilde{a}_{k_-,\vec{k}}|0_{\textrm{SW}}\rangle_F = 0,
    \end{equation}
with $k_- > 0$ and $\vec{k}$ arbitrary. It is straightforward to also show that the final eigenstates $f^F_{k_-,\vec{k}}$ are orthonormal with respect to the Klein-Gordon norm \eqref{eq:KGNorm}.

\section{Rindler observer in a shock wave background}\label{sec:Rindler}

We now consider a probe observer with constant acceleration $a$ in the right Rindler wedge and with its particle horizon located at $u = 0$. There are two distinct cases. The shock wave can be either crossing the accelerated observer ($u_0 < 0$), or the shock wave is hidden beyond the acceleration horizon ($u_0 > 0$). If the shock wave is hidden beyond the acceleration horizon, the global plane wave solutions $f^I_{k_-,\vec{k}}$ will agree with the standard plane wave solutions on the entire right Rindler patch. Moreover, the solution for the scalar field in the right Rindler wedge will coincide with the standard solution without a shock wave and therefore the Bogoliubov coefficients of the accelerated observer with respect to the global solution will be the standard Unruh coefficients. Since there is no trace of the shock wave to be seen in this scenario, the result is trivial. 

We will now focus on the other, more interesting case, where the shock wave is placed at $u_0 < 0$. We consider a constant impact parameter $\vec{x}(\tau) = \vec{b}$ along the worldline. Since we are only interested in the signatures of the supertranslation shift upon crossing of the shockwave, one can place the shock wave infinitesimally close to the particle horizon of the accelerated observer, i.e. at $u_0 = 0^-$. By doing so one avoids unnecessary technical complications due to wave propagation between the shock wave and the particle horizon.

Having defined the setup, the next step is to properly define a basis of modes for both the left and the right Rindler wedges. The solution of the scalar field in the left Rindler wedge can be defined in the standard way that is known from the usual Unruh effect as the shock wave is placed at a $u_0 <0$. Now, since the initial eigenstate shock wave solutions differ from the standard plane wave solutions in the left Rindler patch, the Bogoliubov coefficients $\alpha_L$, $\beta_L$ of the left Rindler observer with respect to the global solution will differ from the standard Unruh coefficients. In particular, the function $f(\vec{x})$ characterizing the shock wave -- or equivalently the supertranslation along $v$ -- will now show up in the Bogoliubov coefficients. We will show in detail how to obtain these modified coefficients in the following. 

\begin{figure}[!htb]
		\centering
		\includegraphics[width=\textwidth]{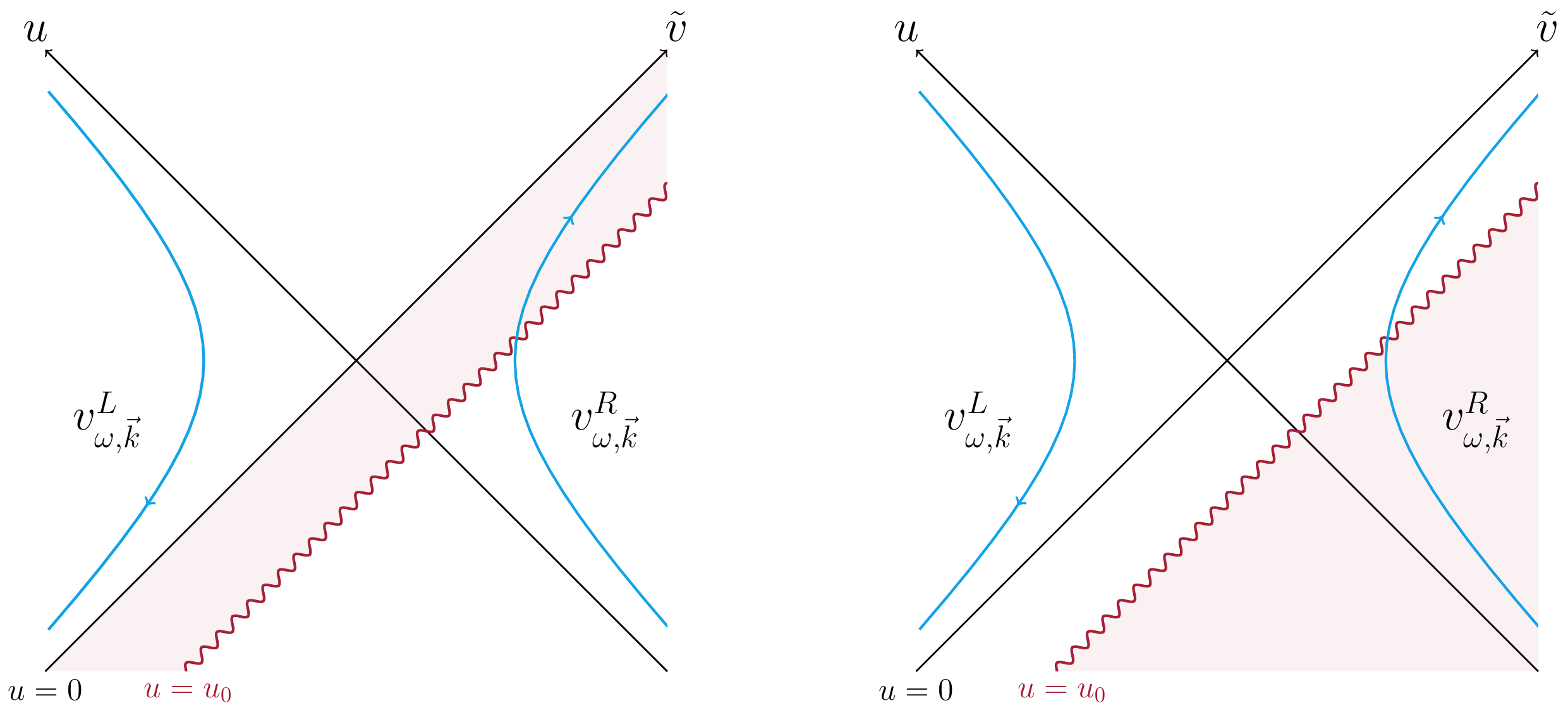}
		\caption{Schematic depiction of the construction of the right Rindler modes. The left picture depicts the case where one starts out with the standard Rindler modes in the region $u<u_0$ and then propagates the solution using the free wave equation for $0>u>u_0$. The right one shows the construction starting with the standard Rindler modes in the region $0>u>u_0$ and then use the free wave equation to propagate the solution backwards for $u<u_0$.}
		\label{fig:Right Rindler}
	\end{figure}

The most obvious way to define a consistent solution in the right Rindler wedge is to proceed exactly as we did previously when constructing the global plane wave solutions in Sec.~\ref{sec:GlobalPlaneWave}. That is, to use the standard result for the solution of the scalar field in the right Rindler wedge up until the location of the shock wave. At $u=u_0$ the solution experiences a supertranslation and after passing the shock wave the solution gets propagated using the free wave equation. What makes this construction subtle is the fact that there are a priori two choices on how to perform this procedure. One is to start with the standard Rindler modes in the region $u<u_0$ and the other is to start with the standard Rindler modes in the region $0>u>u_0$. While the first choice seems to be the most ``natural'' one at first glance it is not obvious how the analytic continuation from the left to the right Rindler patch (that is necessary to compute the Bogoliubov coefficients) works in this case. The second choice on the other hand implies the standard analytic continuation between left and right Rindler modes and, as a consequence, allows the computation of consistent and nontrivial Bogoliubov coefficients. This  is shown in Fig.~\ref{fig:Right Rindler}. We now turn to explaining the details of this construction, following closely the notations of the review \cite{Crispino:2007eb}.

\subsection{Left Rindler wedge}

The Rindler observer in the left patch, i.e. the region $z<-|t|$, is defined via the following coordinates
    \begin{equation}\label{left}
        t=a^{-1}e^{a\bar \xi}\sinh a\bar \tau,\  z=-a^{-1}e^{a\bar \xi}\cosh a\bar \tau, 
    \end{equation}
or
    \begin{equation}
        u=a^{-1}e^{a(\bar\tau+\bar\xi)},\ v=-a^{-1}e^{-a(\bar\tau-\bar\xi)}.
    \end{equation}
The shock wave is outside the left Rindler patch, so the metric is the standard Rindler one 
    \begin{equation}
        \extd s^2=-e^{2a\bar \xi}\left(\extd\bar \tau^2-\extd\bar \xi^2\right)+\sum_{i=2}^{D-1} (\extd x^i)^2.
    \end{equation}
The Klein-Gordon equation in these coordinates takes the form
    \begin{equation}
        \partial^2_{\bar{\tau}}\Phi^{\textrm{L}}=\left[\partial^2_{\bar{\xi}}+e^{2a\bar{\xi}}\Delta\right]\Phi^{\textrm{L}}.
    \end{equation}
Expanding the field in the left wedge in terms of Fourier modes as 
    \begin{equation}
        \Phi^{\textrm{L}}=\int_{0}^{\infty}\extd\omega\int \extd\bm{k} \left(a^L_{\omega,\vec{k}}v^L_{\omega,\vec{k}}+a^{L\dagger}_{\omega,\vec{k}}v^{*L}_{\omega,\vec{k}}\right),
    \end{equation}
one finds that $v^L_{\omega,\vec{k}} = 0$ in the right wedge and in the left wedge is given by \cite{Crispino:2007eb}
    \begin{equation}\label{vL}
        v^L_{\omega,\vec{k}}=\tilde N_\omega K_{\frac{i\omega}{a}}\left(\frac{|\vec{k}|}{a}e^{a\bar\xi}\right)e^{-i\omega\bar\tau+i\vec{k}\cdot\vec{x}},\qquad \tilde N_\omega \equiv \left(\frac{4\sinh\frac{\pi\omega}{a}}{(2\pi)^Da }\right)^{1/2},
    \end{equation}
with $K_{a}(x)$ being the modified Bessel function. Since these are the usual Unruh modes without the presence of a shock wave the coefficients $a^L_{\omega,\vec{k}}$ and $a^{L \dagger}_{\omega,\vec{k}}$ upon quantization also have to satisfy the canonical commutation relations
    \begin{equation}
        [a^L_{\omega,\vec{k}}, a^{L \dagger}_{\omega',\vec{k}'}] = \delta(\omega - \omega') \delta(\vec{k}-\vec{k}').  
    \end{equation}

\subsection{Right Rindler wedge}

Let us now proceed to the more interesting right Rindler wedge $z>|t|$. In the absence of a shock wave, one usually defines the accelerating observer in the right wedge via 
    \begin{equation}
        t=a^{-1}e^{a\xi}\sinh a\tau,\ z=a^{-1}e^{a\xi}\cosh a\tau, 
    \end{equation}
or equivalently 
    \begin{equation}\label{right}
        u=-a^{-1}e^{-a(\tau-\xi)},\ v=a^{-1}e^{a(\tau+\xi)}.
    \end{equation}
Now, in the presence of the shock wave, one needs to use the shifted coordinates $\tilde u = u$, \mbox{$\tilde{v} = v - f(\vec{b})\Theta(u-u_0)$} in the right Rindler wedge so that the trajectory of the accelerated observer is continuous. Even though the trajectory of the accelerated observer in these coordinates is the standard Rindler one
    \begin{equation}\label{eq:RindlerObserver}
        (u(\tau), \tilde v(\tau)) = \left(-\frac{e^{-a \tau}}{a}, \frac{e^{a \tau}}{a}\right),
    \end{equation}
the metric differs from the standard Rindler metric outside the worldline and is given by
    \begin{equation}
        ds^2= -\extd u \extd\tilde v + \left(f(\vec{x})-f(\vec{b})\right) \delta(u - u_0) \extd u^2 + \sum_{i=2}^{D-1} (\extd x^i)^2 .
    \end{equation}
It is straightforward to check that the observer has constant acceleration $a > 0$ along the worldline described by \eqref{eq:RindlerObserver}. This is the same coordinate system in which null geodesics at constant impact parameter $\vec{x} = \vec{b}$ are continuous. In right Rindler-type coordinates $u = -a^{-1}e^{-a (\tau - \xi)}$, $\tilde v = a^{-1}e^{a (\tau +\xi)}$, one has 
    \begin{equation}
        \extd s^2 = e^{2 a \xi} \left(- \extd\tau^2 + \extd\xi^2\right) + \sum_{i=2}^{D-1} (\extd x^i)^2
        + \left(f(\vec{x})-f(\vec{b})\right) \delta(u - u_0) e^{-2 a (\tau - \xi)} \left(\extd\tau - \extd\xi\right)^2.
    \end{equation}
Now, the metric is $\tau$ dependent, so the right Rindler modes will have a more complicated form than in the left wedge.

In order to ensure continuity between the left and right Rindler modes at the bifurcation point of the particle horizon we construct the right Rindler modes as follows. In between the region of the future particle horizon at $u=0$ and the shock wave at $u = u_0$, the positive frequency modes $v^R_{\omega,\vec{k}}(\tau, \zeta,\vec{x})$ can be straightforwardly obtained by the expressions for the modes in the left wedge simply by replacing $\bar{\tau}\rightarrow\tau$ and $\bar{\xi}\rightarrow\xi$
    \begin{equation}\label{vR}
        v^R_{\omega,\vec{k}}(\tau, \zeta,\vec{x})\big|_{0>u>u_0}=\tilde N_\omega K_{\frac{i\omega}{a}}\left(\frac{|\vec{k}|}{a}e^{a\xi}\right)e^{-i\omega\tau+i\vec{k}\cdot\vec{x}},\qquad \tilde N_\omega \equiv \left(\frac{4\sinh\frac{\pi\omega}{a}}{(2\pi)^D a}\right)^{1/2}.
    \end{equation}
The presence of the shock wave implies that -- in complete analogy to the global Minkowski modes that were introduced previously in Section~\ref{sec:GlobalPlaneWave} -- the modes at $u \rightarrow u_0+0^+$ are related to the modes at $u \rightarrow u_0+0^-$ by a ``planar supertranslation'' of the form 
    \begin{equation}
        \tilde v \rightarrow \tilde v - f(\vec{x})\qquad \text{or} \qquad  v \rightarrow v - (f(\vec{x}) - f(\vec{b})). 
    \end{equation}
Thus, the right Rindler modes right after passing the shock wave at $u_0+0^-$ are given by
    \begin{equation}
        v^R_{\omega,\vec{k}}(\tau, \zeta,\vec{x})\big|_{u=u_0+0^-}=\tilde N_\omega K_{\frac{i\omega}{a}}\left(\frac{|\vec{k}|}{a}e^{a\xi_>}\right)e^{-i\omega\tau_>+i\vec{k}\cdot\vec{x}},\qquad \tilde N_\omega \equiv \left(\frac{4\sinh\frac{\pi\omega}{a}}{(2\pi)^D a}\right)^{1/2},
    \end{equation}
where 
    \begin{subequations}
        \begin{align}
            \xi_> &= \xi + \frac{1}{2a}\log\left[1-v^{-1} \left(f(\vec{x})-f(\vec{b})\right)\right],\\
            \tau_> &= \tau + \frac{1}{2a}\log\left[1-v^{-1} \left(f(\vec{x})-f(\vec{b})\right)\right].
        \end{align}
    \end{subequations}
The modes in the region $u<u_0$ are then propagated using the free wave equation in right Rindler coordinates. This procedure leads to the analogue of Section  \ref{sec:GlobalPlaneWave} in right Rindler coordinates. However, as we show in the next section we will only need explicitly the expression for $v^R_{\omega,\vec{k}}$ in the region $0>u>u_0$ in order to compute the Bogoliubov coefficients.

Using these modes the scalar field in the right Rindler wedge can then be Fourier expanded as 
    \begin{equation}
    \Phi^{\textrm{R}}=\int_{0}^{\infty}\extd\omega\int \extd\bm{k} \left(a^R_{\omega,\vec{k}}v^R_{\omega,\vec{k}}+a^{R\dagger}_{\omega,\vec{k}}v^{*R}_{\omega,\vec{k}}\right).
    \end{equation}

\subsection{Rindler shock wave state}

Using the left and right modes we constructed previously one can express a general solution as 
    \begin{equation}
        \Phi=\int_0^{\infty}\extd\omega \int \extd\bm{k}\left(a^R_{\omega,\vec{k}}v^R_{\omega,\vec{k}}+a^{R\dagger}_{\omega,\vec{k}}v^{*R}_{\omega,\vec{k}} +a^L_{\omega,\vec{k}}v^L_{\omega,\vec{k}}+a^{L\dagger}_{\omega,\vec{k}}v^{*L}_{\omega,\vec{k}}\right).\label{dec}
    \end{equation}
Since $v^R_{\omega,\vec{k}}$ and $v^L_{\omega,\vec{k}}$ are positive frequency modes the vacuum state for the Rindler observers in the presence of the shock wave is defined by
    \begin{equation}
        a^R_{\omega,\vec{k}}|0_{\textrm{Rin}} \rangle =a^L_{\omega,\vec{k}}|0_{\textrm{Rin}} \rangle =0.
    \end{equation}
In order to determine a possible influence of the function $f(\vec{x})$ on the spectrum of the Unruh temperature that the accelerated observer measures one has to compute the Bogoliubov coefficients that relate the left and right Rindler modes $a^L_{\omega,\vec{k}}$ and $a^R_{\omega,\vec{k}}$ with the global plane wave modes $f^{I}_{k_-,\vec{k}}$. This will be done in the following section. 

\section{Bogoliubov transform between Rindler and global vacua}
\label{sec:Bogoliubov}

In this section we compute the Bogoliubov coefficients that provide the map between the global plane wave modes and the Rindler modes. We first outline the computation starting with the global vacuum defined from the initial eigenstate plane wave modes $f^I_{k_-,\vec{k}}$. As a cross-check we then repeat the calculation starting with the global vacuum defined with respect to the final eigenstate modes $f^F_{k_-,\vec{k}}$. We find that either choice leads to a consistent set of Bogoliubov coefficients and also to the same Unruh spectrum.

In order to keep the computations in the following as simple as possible we set, without loss of generality,  $f(\vec{b}) = 0$. The reader can restore $f(\vec{b})$ by shifting $f(\vec{x})$ appropriately in the upcoming expressions. This means that there will also be no distinctions between $\tilde v$ and $v$. We will also set the total spacetime dimension to be $D = 4$, which is the physically relevant case. The appropriate prefactors for dimensions higher than four can be restored in a  straightforward manner.

\subsection{Initial eigenstate global vacuum}

Since, the left and right Rindler modes $a^L_{\omega,\vec{k}}$ and $a^R_{\omega,\vec{k}}$ as well as the global plane wave modes $f^{I}_{k_-,\vec{k}}$ form a complete set of modes one can express either set in terms of the other. 

In particular,
    \begin{subequations}
    \begin{align}\label{eq:RindlerModes}
        v^R_{\omega,\vec{k}}(\tau,\xi,\vec{x})&=\int_{-\infty}^{\infty}\extd l_z\int \extd\bm{l}\left(\alpha^R_{\omega,\vec{k};l_-,\vec{l}}\,f^I_{l_-,\vec{l}}\,(u,v,\vec{x})+\beta^R_{\omega,\vec{k};l_-,\vec{l}}\,f^{I*}_{l_-,\vec{l}}\,(u,v,\vec{x})\right),\\
        v^L_{\omega,\vec{k}}(\bar{\tau},\bar{\xi},\vec{x})&=\int_{-\infty}^{\infty}\extd l_z\int \extd\bm{l}\left(\alpha^L_{\omega,\vec{k};l_-,\vec{l}}\,f^I_{l_-,\vec{l}}\,(u,v,\vec{x})+\beta^L_{\omega,\vec{k};l_-,\vec{l}}\,f^{I*}_{l_-,\vec{l}}\,(u,v,\vec{x})\right).\label{vRvL}
    \end{align}
    \end{subequations}
The most straightforward way to obtain the Bogoliubov coefficients $\alpha^{L/R}_{\omega,\vec{k};l_-,\vec{l}}$ and $\beta^{L/R}_{\omega,\vec{k};l_-,\vec{l}}$ is to use the Klein-Gordon norm \eqref{eq:KGNorm}. The Rindler modes $v^{L/R}_{\omega,\vec{k}}(\bar{\tau},\bar{\xi},\vec{x})$ are quite complicated and, as such, evaluating the Klein-Gordon norm involves quite a bit of algebra. However, near the (future) particle horizon these modes simplify substantially and allow for a very efficient computation. This simplification is also explained in detail in \cite{Crispino:2007eb, Takagi:1986kn}.

We start first with obtaining the Bogoliubov coefficients for the right Rindler wedge. The particle horizon $\mathcal H^+$ is located at $u=0$ and $t > 0$ or $\xi \rightarrow -\infty$. In the limit $u\rightarrow0$ and for $t>0$ we have 
    \begin{equation}\label{vr2}
        v^R_{\omega,\vec{k}}(\tau,\xi,\vec{x})\sim  \int_{-\infty}^{\infty}\extd l_z\int \extd\bm{l}\left(\alpha^R_{\omega,\vec{k};l_-,\vec{l}}\,f^I_{l_-,\vec{l}}\,(0,v,\vec{x})+\beta^R_{\omega,\vec{k};l_-,\vec{l}}\,f^{I*}_{l_-,\vec{l}}\,(0,v,\vec{x})\right).
    \end{equation}
Now, using
 the small function argument of the Bessel function, see appendix A of \cite{Crispino:2007eb}, one obtains at the horizon
    \begin{equation}\label{vRa}
        v^R_{\omega,\vec{k}}\sim \frac{i}{4\pi \sqrt{a \sinh \frac{\pi\omega}{a}}}e^{i \vec{k} \cdot \vec{x}} \left( \frac{\left(\frac{|k|}{2a}\right)^{\frac{i\omega}{a}}e^{- i \omega (\tau - \xi )}}{\Gamma (1+\frac{i \omega}{a})}-\frac{\left(\frac{|k|}{2a}\right)^{-\frac{i\omega}{a}}e^{- i \omega (\tau + \xi )}}{\Gamma (1-\frac{i \omega}{a})}\right). 
    \end{equation}
The first term oscillates infinitely many times as one approaches $\mathcal{H}^+$ at $u=0$ since $u\propto e^{a(\xi-\tau)}\rightarrow0$ and is bounded. Hence, it should be regarded as zero \cite{Crispino:2007eb} and we will ignore it in the following. Performing the inverse Fourier transform using \eqref{vr2} and \eqref{vRa} one obtains 
    \begin{equation}\label{eq:BogoliubovRight1}
        \alpha^R_{\omega,\vec{k};l_-,\vec{l}}=\frac{e^{\frac{\pi\omega}{2a}-i\frac{\vec{k}^2}{4l_-}u_0}}{\sqrt{4\pi a l_t\sinh\frac{\pi\omega}{a}}}\left(\frac{|k|}{2l_-}\right)^{-i\omega/a}\phi_{-l_-}(\vec{k}-\vec{l}),
    \end{equation}
where the mixing factor  $\phi_{l_-}(\vec{k}-\vec{l})$ has been defined previously in \eqref{eq:PhiDef}. The other coefficients $\beta^R_{\omega,\vec{k};l_-,\vec{l}}$ can be obtained in the same manner and one obtains the closely related expression
    \begin{equation}\label{eq:BogoliubovRight2}
         \beta^R_{\omega,\vec{k};l_-,\vec{l}}=-\frac{e^{-\frac{\pi\omega}{2a}+i\frac{\vec{k}^2}{4l_-}u_0}}{\sqrt{4\pi a l_t\sinh\frac{\pi\omega}{a}}}\left(\frac{|k|}{2l_-}\right)^{-i\omega/a}\phi_{l_-}(\vec{k}+\vec{l}).
    \end{equation}
The Bogoliubov coefficients for the left Rindler wedge can be computed using almost the same reasoning as before. For the left Rindler modes we have to approach the particle horizon $\mathcal{H}^-$ with $u=0$ and $t<0$. In this limit the Rindler modes expressed in terms of the global plane wave modes \eqref{vRvL} take the form
    \begin{equation}\label{eq:LeftRindlerHorizon}
        v^L_{\omega,\vec{k}}(\tau,\xi,\vec{x})\sim  \int_{-\infty}^{\infty}\extd l_z  \int \extd\bm{l} \,\left(\alpha^L_{\omega,\vec{k};l_-,\vec{l}}\,f^I_{l_-,\vec{l}}\,(0,v,\vec{x})+\beta^L_{\omega,\vec{k};l_-,\vec{l}}\,f^{I*}_{l_-,\vec{l}}\,(0,v,\vec{x})\right).
    \end{equation}
The near horizon expansion of the solution of the left Rindler modes expressed in Rindler coordinates \eqref{vL} in this limit is given by
    \begin{equation}\label{eq:LeftRindlerHorizonExpansion}
        v^L_{\omega,\vec{k}}\sim \frac{i}{4\pi \sqrt{a \sinh \frac{\pi\omega}{a}}}e^{i \vec{k} \cdot \vec{x}} \left( \frac{\left(\frac{|k|}{2a}\right)^{\frac{i\omega}{a}}e^{- i \omega (\bar{\tau} - \bar{\xi})}}{\Gamma (1+\frac{i \omega}{a})}-\frac{\left(\frac{|k|}{2a}\right)^{-\frac{i\omega}{a}}e^{- i \omega (\bar{\tau} + \bar{\xi})}}{\Gamma (1-\frac{i \omega}{a})}\right).
    \end{equation}
In terms of the left Rindler coordinates $\bar{\xi}$ and $\bar{\tau}$ the location of the horizon at $u=0$ indicates via $u\propto e^{a(\bar{\xi}+\bar{\tau})}\rightarrow0$ that the second term now oscillates infinitely many times, while being bounded and thus has to be discarded. Using again the inverse Fourier transform of \eqref{eq:LeftRindlerHorizon} the Bogoliubov coefficients for the left Rindler wedge read
    \begin{subequations}\label{eq:BogoliubovLeft}
        \begin{align}
             \alpha^L_{\omega,\vec{k};l_-,\vec{l}}&=\frac{e^{\frac{\pi\omega}{2a}-i\frac{\vec{k}^2}{4l_-}u_0}}{\sqrt{4\pi a l_t\sinh\frac{\pi\omega}{a}}}\left(\frac{|k|}{2l_-}\right)^{i\omega/a}\phi_{-l_-}(\vec{k}-\vec{l}),\\
            \beta^L_{\omega,\vec{k};l_-,\vec{l}}&=-\frac{e^{-\frac{\pi\omega}{2a}+i\frac{\vec{k}^2}{4l_-}u_0}}{\sqrt{4\pi a l_t\sinh\frac{\pi\omega}{a}}}\left(\frac{|k|}{2l_-}\right)^{i\omega/a}\phi_{l_-}(\vec{k}+\vec{l}).
        \end{align}
     \end{subequations}
This computation shows that the shock wave has a nontrivial effect on the Bogoliubov coefficients relating the shock wave vacuum to the Rindler vacuum. 

After quantization, the creation and annihilation operators $a^R_{\omega,\vec{k}}$, $a^R_{\omega,\vec{k}}$, $a^L_{\omega,\vec{k}}$ and $a^L_{\omega,\vec{k}}$ have to satisfy the canonical commutation relations
    \begin{equation}
        [a^R_{\omega,\vec{k}},a^{R\dagger}_{\omega',\vec{k}'}]  = [a^L_{\omega,\vec{k}},a^{L\dagger}_{\omega',\vec{k}'}]    = \delta(\omega-\omega')\delta(\vec{k}-\vec{k}').
    \end{equation}
This, in turn also puts restrictions on the Bogoliubov coefficients
    \begin{subequations}\label{eq:BogoliubovConsistency1}
        \begin{align}
            	\int \extd l_z \extd\bm{l}\left(\alpha^{L/R*}_{\omega,\vec{k};l_-,\vec{l}}\,\alpha^{L/R}_{\omega',\vec{k}';l_-,\vec{l}}-\beta^{L/R*}_{\omega,\vec{k};l_-,\vec{l}}\,\beta^{L/R}_{\omega',\vec{k}';l_-,\vec{l}}\right)  &= \delta(\omega-\omega')\delta(\vec{k}-\vec{k}'),\label{eq:BogoliubovConsistency1a}\\
                \int \extd l_z \extd\bm{l}\left(\alpha^{L/R}_{\omega,\vec{k};l_-,\vec{l}}\,\beta^{L/R}_{\omega',\vec{k}';l_-,\vec{l}}-\beta^{L/R}_{\omega,\vec{k};l_-,\vec{l}}\,\alpha^{L/R}_{\omega',\vec{k}';l_-,\vec{l}}\right)  &= 0.
        \end{align}
    \end{subequations}
On the other hand, from
    \begin{equation}
        [a_{l_-,\vec{l}},a^\dagger_{l_-',\vec{l}'}]=\delta(l_z-l_z')\delta(\vec{l}-\vec{l}'),\quad [a_{l_-,\vec{l}},a_{l_-',\vec{l}'}]=[a^\dagger_{l_-,\vec{l}},a^\dagger_{l_-',\vec{l}'}]=0,
    \end{equation}
one can infer that
    \begin{subequations}\label{eq:BogoliubovConsistency2}
        \begin{align}
            \sum_{A=L,R}\int \extd\omega \extd\bm{k}\left(\alpha^{A}_{\omega,\vec{k};l_-,\vec{l}}\,\alpha^{A*}_{\omega,\vec{k};l_-',\vec{l}'}-\beta^{A*}_{\omega,\vec{k};l_-,\vec{l}}\,\beta^{A}_{\omega,\vec{k};l_-',\vec{l}'}\right) & =  \delta(l_z-l_z')\delta(\vec{l}-\vec{l}'),\label{cons3}\\
	       \sum_{A=L,R} \int \extd\omega \extd\bm{k}\left(\alpha^{A}_{\omega,\vec{k};l_-,\vec{l}}\,\beta^{A*}_{\omega,\vec{k};l_-',\vec{l}'}-\beta^{A*}_{\omega,\vec{k};l_-,\vec{l}}\,\alpha^{A}_{\omega,\vec{k};l_-',\vec{l}'}\right) & =  0,
        \end{align}
    \end{subequations}
has to hold as well. The relations \eqref{eq:BogoliubovConsistency1} and \eqref{eq:BogoliubovConsistency2} serve as consistency checks that ought to be obeyed by sensible Bogoliubov coefficients. It is straightforward using \eqref{propphi} to check that the  coefficients \eqref{eq:BogoliubovRight1}, \eqref{eq:BogoliubovRight2} and \eqref{eq:BogoliubovLeft} do obey these consistency conditions.

\subsection{Final eigenstate global vacuum}

The preceding arguments can be also applied to compute the Bogoliubov coefficients between the final eigenstate solutions $f^F_{k_-,\vec{k}}$ and the Rindler modes $v^{R/L}_{\omega,\vec{k}}$. Now the Rindler modes are expanded as
    \begin{subequations}
    \begin{align}\label{eq:RindlerModesFinal}
        v^R_{\omega,\vec{k}}(\tau,\xi,\vec{x})&=\int_{-\infty}^{\infty}\extd l_z\int \extd\bm{l}\left(\tilde{\alpha}^R_{\omega,\vec{k};l_-,\vec{l}}\,f^F_{l_-,\vec{l}}\,(u,v,\vec{x})+\tilde{\beta}^R_{\omega,\vec{k};l_-,\vec{l}}\,f^{F*}_{l_-,\vec{l}}\,(u,v,\vec{x})\right),\\
        v^L_{\omega,\vec{k}}(\bar{\tau},\bar{\xi},\vec{x})&=\int_{-\infty}^{\infty}\extd l_z\int \extd\bm{l}\left(\tilde{\alpha}^L_{\omega,\vec{k};l_-,\vec{l}}\,f^F_{l_-,\vec{l}}\,(u,v,\vec{x})+\tilde{\beta}^L_{\omega,\vec{k};l_-,\vec{l}}\,f^{F*}_{l_-,\vec{l}}\,(u,v,\vec{x})\right).
    \end{align}
    \end{subequations}
At $\mathcal{H}^{-/+}$ this expansion reduces to
    \begin{equation}
        v^{L/R}_{\omega,\vec{k}}(\tau,\xi,\vec{x})\sim  \int_{-\infty}^{\infty}\extd l_z \int \extd\bm{l} \left(\tilde{\alpha}^{L/R}_{\omega,\vec{k};l_-,\vec{l}}\,f^F_{l_-,\vec{l}}\,(0,v,\vec{x})+\tilde{\beta}^{L/R}_{\omega,\vec{k};l_-,\vec{l}}\,f^{F*}_{l_-,\vec{l}}\,(0,v,\vec{x})\right).
    \end{equation}
The near horizon expansions of $v^{L/R}_{\omega,\vec{k}}$ are not affected by the choice of either initial or final eigenstate solution as global vacuum and thus are still given by \eqref{vRa} and \eqref{eq:LeftRindlerHorizonExpansion}. As a result the Bogoliubov coefficients are again straightforward to compute by a inverse Fourier transformation and lead to
    \begin{subequations}\label{eq:BogoliubovLeftFinal}
        \begin{align}
             \tilde{\alpha}^L_{\omega,\vec{k};l_-,\vec{l}}&=\frac{e^{\frac{\pi\omega}{2a}-i\frac{\vec{k}^2}{4l_-}u_0}}{\sqrt{4\pi a l_t\sinh\frac{\pi\omega}{a}}}\left(\frac{|k|}{2l_-}\right)^{i\omega/a}\delta(\vec{k}-\vec{l}),\\
            \tilde{\beta}^L_{\omega,\vec{k};l_-,\vec{l}}&=-\frac{e^{-\frac{\pi\omega}{2a}+i\frac{\vec{k}^2}{4l_-}u_0}}{\sqrt{4\pi a l_t\sinh\frac{\pi\omega}{a}}}\left(\frac{|k|}{2l_-}\right)^{i\omega/a}\delta(\vec{k}+\vec{l}),
        \end{align}
     \end{subequations}    
and
    \begin{subequations}\label{eq:BogoliubovRightFinal}
        \begin{align}
             \tilde{\alpha}^R_{\omega,\vec{k};l_-,\vec{l}}&=\frac{e^{\frac{\pi\omega}{2a}-i\frac{\vec{k}^2}{4l_-}u_0}}{\sqrt{4\pi a l_t\sinh\frac{\pi\omega}{a}}}\left(\frac{|k|}{2l_-}\right)^{-i\omega/a}\delta(\vec{k}-\vec{l}),\\
            \tilde{\beta}^R_{\omega,\vec{k};l_-,\vec{l}}&=-\frac{e^{-\frac{\pi\omega}{2a}+i\frac{\vec{k}^2}{4l_-}u_0}}{\sqrt{4\pi a l_t\sinh\frac{\pi\omega}{a}}}\left(\frac{|k|}{2l_-}\right)^{-i\omega/a}\delta(\vec{k}+\vec{l}).
        \end{align}
     \end{subequations}
These Bogoliubov coefficients are essentially the same as in the original setup considered by Unruh in pure Minkowski space \cite{Unruh:1976db}, up to a phase. As a consequence, the radiation spectrum will be the standard one without any traces of the supertranslation induced by the shock wave. In the following, we will show that this is, indeed, the case starting from both global plane wave vacua.

\subsection{Unruh effect and thermal density matrix}

Having at our disposal a consistent set of Bogoliubov coefficients that relate the creation and annihilation operators of the Rindler vacuum to the initial or final global vacuum, the next step is to compute the expectation value of the left and right Rindler number operator \mbox{$\textrm{N}_{\omega,\vec{k}}^{L/R}=a^{L/R\dagger}_{\omega,\vec{k}}a^{L/R}_{\omega,\vec{k}}$} acting on either the initial or final shock wave vacuum state. This expectation value can be directly computed using the Bogoliubov coefficients $\beta^{L/R}_{\omega,\vec{k};l_-,\vec{l}}\,$, see e.g. \cite{Crispino:2007eb}. For the shock wave vacuum corresponding to the initial eigenstate solution this expectation value for the left and right Rindler modes is given by
    \begin{equation}
        {}_I\langle 0_{SW} | \textrm{N}^{L/R}_{\omega, \vec{k}} | 0_{SW} \rangle_I = \int \extd l_z \int \extd \bm{l}\; \Big|\beta^{L/R}_{\omega,\vec{k};l_-,\vec{l}}\,\Big|^2.
    \end{equation}
where the Bogoliubov coefficients are given by  \eqref{eq:BogoliubovLeft}, \eqref{eq:BogoliubovRight1} and \eqref{eq:BogoliubovRight2}. Making use of \eqref{propphi}, it is easy to show that these coefficients satisfy
    \begin{equation}
        \int\extd\bm{l}\Big|\alpha^{L/R}_{\omega,\vec{k};l_-,\vec{l}}\Big|^2= \int\extd\bm{l}\,e^{\frac{2\pi\omega}{a}}\Big|\beta^{L/R}_{\omega,\vec{k};l_-,\vec{l}}\Big|^2.
    \end{equation}
 
Then one can write \eqref{eq:BogoliubovConsistency1a} for $\omega=\omega'$ and $\vec{k}=\vec{k}'$ as
    \begin{equation}
        \int \extd l_z \extd\bm{l}\left(e^{\frac{2\pi\omega}{a}}-1\right) \Big|\beta^{L/R}_{\omega,\vec{k};l_-,\vec{l}}\Big|^2= \delta(0)\delta(\vec{0}).
    \end{equation}
The singular delta functions can be regulated using wave packets as usual. From this it directly follows that the expectation value of the left and right Rindler number operators are the same as in the standard Unruh case, i.e.
    \begin{equation}
        {}_I\langle 0_{SW} | \textrm{N}^{L/R}_{\omega, \vec{k}} | 0_{SW} \rangle_I = \int \extd l_z \int \extd \bm{l}\; \Big|\beta^{L/R}_{\omega,\vec{k};l_-,\vec{l}}\,\Big|^2=\frac{\delta(0)\delta(\vec{0})}{e^{\frac{2\pi\omega}{a}}-1}.\label{UnruhRad}
    \end{equation}
For the future eigenstate vacuum $|0_{SW}\rangle_F$ one obtains the exact same result. This shows that a Rindler observer in a shock wave background measures precisely the same Unruh temperature as in the original, pure Minkowski setup. 


To show that shock wave vacuum restricted to left/right Rindler wedge is indeed a thermal state, we should find the reduced density matrix in left/right Rindler wedge. 
We observe that the standard relations 
 \be
 \beta^L_{\omega,\vec{k};l_-\vec{l}}=-e^{-\pi\omega/a}\alpha^{R*}_{\omega,-\vec{k};l_-\vec{l}}\ ,\qquad \qquad  \beta^R_{\omega,\vec{k};l_-\vec{l}}=-e^{-\pi\omega/a}\alpha^{L*}_{\omega,-\vec{k};l_-\vec{l}}\ 
 \ee 
 are still obeyed even in the presence of the supertranslation factor in the case of the initial quantization. Following the derivation of Unruh \cite{Unruh:1976db} we deduce that the shock wave vacuum  is annihilated by the operators 
 \bea
(a^L_{\omega,\vec{k}}-e^{-\pi\omega/a} a^{R\dagger}_{\omega,-\vec{k}})|0_{SW}\rangle&=&0,\\
(a^R_{\omega,\vec{k}}-e^{-\pi\omega/a} a^{L\dagger}_{\omega,-\vec{k}})|0_{SW}\rangle&=&0.
 \eea 
 This implies that the density of left Rindler modes at each $\omega$, $\vec{k}$ is the same as the density of right Rindler modes at each $\omega$, $-\vec{k}$ ,
 \be
( a^{R\dagger}_{\omega,\vec{k}}a^R_{\omega,\vec{k}}-a^{L\dagger}_{\omega,-\vec{k}}a^L_{\omega,-\vec{k}})|0_{SW}\rangle=0.
 \ee 
Following \cite{Crispino:2007eb}, the shock wave vacuum can be rewritten as
\bea
|0_{SW}\rangle &=& \prod_{\omega,\vec{k}} C_{\omega} e^{\zeta_{\omega} a^{R\dagger}_{\omega,\vec{k}}a^{L\dagger}_{\omega,-\vec{k}}}|0_R\rangle,\\
&=& \prod_{\omega,\vec{k}}\Big( C_{\omega} \sum_{n(\omega,\vec{k})=0}^{\infty} (\zeta_\omega)^{n(\omega,\vec{k})} |n(\omega,-\vec{k}),L\rangle \otimes |n(\omega,\vec{k}),R\rangle\Big)
\eea
 where $\zeta_{\omega}=e^{-\pi\omega/a}$, $C_{\omega}=\sqrt{1-\zeta_{\omega}^2}$ and $n(\omega,\vec{k})$ is the density of states defined from 
\bea
 |n(\omega,-\vec{k}),L\rangle \otimes |n(\omega,\vec{k}),R\rangle = \frac{1}{n(\omega,\vec{k})!} \left(a^{R\dagger}_{\omega,\vec{k}}a^{L\dagger}_{\omega,-\vec{k}} \right)^{n(\omega,\vec{k})} |0_R\rangle.
\eea 
After taking a trace, the reduced density matrix in the left/right Rindler wedge is 
\bea
{\rho}_{L}&=&\prod_{\omega,\vec{k}}\Big(C_{\omega}^2\sum_{n(\omega,\vec{k})=0}^{\infty}(\zeta_\omega)^{2n(\omega,\vec{k})} |n(\omega,-\vec{k}),L\rangle\langle n(\omega,-\vec{k}),L|\Big),\\
{\rho}_{R}&=&\prod_{\omega,\vec{k}}\Big(C_{\omega}^2\sum_{n(\omega,\vec{k})=0}^{\infty} (\zeta_\omega)^{2n(\omega,\vec{k})} |n(\omega,\vec{k}),R\rangle\langle n(\omega,\vec{k}),R|\Big).
\eea 
This proves that the shock wave vacuum restricted to left/right wedge is a thermal state with temperature $T=\frac{a}{2\pi}$.

\subsection{Particle number in global one-particle states}
\label{onep}

Instead of considering the initial or final global vacuum, let us now consider a one-particle state in either choice of quantization scheme. We denote these states as 
\bea
|p_-, \vec{p} \rangle_I \equiv a^\dagger_{p_-,\vec{p}} | 0_{SW} \rangle_I , \qquad |p_- ,\vec{p} \rangle_F \equiv \tilde{a}^\dagger_{p_-,\vec{p}} | 0_{SW} \rangle_F . 
\eea
Since the initial (resp. final) global vacuum is a defined from momentum eigenvalues at $u < u_0$ (resp. $u > u_0$), these two one-particle states for a given momentum $(p_-,\vec{p})$ are distinct. 

What is the particle number as seen by a Rindler observer? We would like to obtain
\bea
  {}_{I/F}\langle p_-, \vec{p} | \textrm{N}^{L/R}_{\omega, \vec{k}} |p_-, \vec{p} \rangle_{I/F}.
\eea
The annihilation operators in the left/right Rindler wedge can be related to either the initial or final global shock wave creation and annihilation operators via
    \begin{equation}
            a^{L/R}_{\omega,\vec{k}}  =\int\extd l_z\extd\bm{l}\left(\alpha^{L/R*}_{\omega, \vec{k} ;l_- ,\vec{l}}a_{l_-,\vec{l}}-\beta^{L/R*}_{\omega, \vec{k} ;l_- ,\vec{l}}a^\dagger_{l_-,\vec{l}}\right)
            =\int\extd l_z\extd\bm{l}\left(\tilde{\alpha}^{L/R*}_{\omega, \vec{k} ;l_- ,\vec{l}}\tilde a_{l_-,\vec{l}}-\tilde{\beta}^{L/R*}_{\omega, \vec{k} ;l_- ,\vec{l}}\tilde a^\dagger_{l_-,\vec{l}}\right).
    \end{equation}
Using the canonical commutation relations
    \begin{equation}
        [a_{p_-,\vec{p}},a^{\dagger}_{q_-,\vec{q}}]  = [\tilde{a}_{p_-,\vec{p}},\tilde{a}^{\dagger}_{q_-,\vec{q}}]    = \delta(p_z-q_z)\delta(\vec{p}-\vec{q}),
    \end{equation}
one obtains after a straightforward calculation
    \begin{subequations}
        \begin{align}
            {}_{I}\langle p_-, \vec{p} | \textrm{N}^{L/R}_{\omega, \vec{k}} |p_-, \vec{p} \rangle_{I} & = \Big|\alpha^{L/R}_{\omega,\vec{k};p_-,\vec{p}}\,\Big|^2+\Big|\beta^{L/R}_{\omega,\vec{k};p_-,\vec{p}}\,\Big|^2 + \langle p_-,\vec{p}|p_-,\vec{p}\rangle\int \extd l_z \int \extd \bm{l}\; \Big|\beta^{L/R}_{\omega,\vec{k};l_-,\vec{l}}\,\Big|^2,\\
            {}_{F}\langle p_-, \vec{p} | \textrm{N}^{L/R}_{\omega, \vec{k}} |p_-, \vec{p} \rangle_{F} & = \Big|\tilde{\alpha}^{L/R}_{\omega,\vec{k};p_-,\vec{p}}\,\Big|^2+\Big|\tilde{\beta}^{L/R}_{\omega,\vec{k};p_-,\vec{p}}\,\Big|^2 + \langle p_-,\vec{p}|p_-,\vec{p}\rangle\int \extd l_z \int \extd \bm{l}\; \Big|\tilde{\beta}^{L/R}_{\omega,\vec{k};l_-,\vec{l}}\,\Big|^2.
        \end{align}
    \end{subequations}
Inserting the expressions for the Bogoliubov coefficients \eqref{eq:BogoliubovRight1}, \eqref{eq:BogoliubovRight2} and \eqref{eq:BogoliubovRightFinal} derived previously one finds
    \begin{subequations}
        \begin{align}
            {}_{I}\langle p_-, \vec{p} | \textrm{N}^{L/R}_{\omega, \vec{k}} |p_-, \vec{p} \rangle_{I} & = \frac{1}{2 \pi a p_t}\left(\frac{\Big|\phi_{-p_-}(\vec{k}-\vec{p})\Big|^2}{1-e^{-\frac{2\pi\omega}{a}}}+\frac{\Big|\phi_{p_-}(\vec{k}+\vec{p})\Big|^2}{e^{\frac{2\pi\omega}{a}}-1}\right) + \langle p_-,\vec{p}|p_-,\vec{p}\rangle_I \frac{\delta(0)\delta(\vec{0})}{e^{\frac{2\pi\omega}{a}}-1},\\
            {}_{F}\langle p_-, \vec{p} | \textrm{N}^{L/R}_{\omega, \vec{k}} |p_-, \vec{p} \rangle_{F} & = \frac{1}{2 \pi a p_t}\left(\frac{\Big|\delta(\vec{k}-\vec{p})\Big|^2}{1-e^{-\frac{2\pi\omega}{a}}}+\frac{\Big|\delta(\vec{k}+\vec{p})\Big|^2}{e^{\frac{2\pi\omega}{a}}-1}\right) + \langle p_-,\vec{p}|p_-,\vec{p}\rangle_F\frac{\delta(0)\delta(\vec{0})}{e^{\frac{2\pi\omega}{a}}-1}.
        \end{align}
    \end{subequations}
The last term in both expressions encodes the vacuum expectation value found previously, which is the same for both choices of quantization. It is independent of the incoming momentum and therefore encodes the spontaneous Unruh emission. The first two terms, however, differ. The one-particle states defined in the initial eigenstate quantization scheme have an occupation number that depends upon the shock wave form factor.  We conclude that while the vacuum expectation value for the number operator in the right Rindler wedge coincides in both quantization schemes the occupation number differs for excited states.  This difference leads to distinct properties of stimulated emission and scattering.

We can generalize our previous considerations to one-particle superposed states. We denote such states as
\be
|\gamma\rangle_I \equiv\int \extd p_z \int\extd\bm{p}\ \gamma_{p_-,\vec{p}}\ |p_-,\vec{p}\rangle_I
\ee 
where we select the initial eigenstate quantization. The norm of the state is normalized to 1 so 
\be
\int\extd p_z \int\extd\bm{p} \ |\gamma_{p_-,\vec{p}}|^2=1.
\ee 
One finds 
\be\label{supp}
{}_I\langle\gamma |\textrm{N}^{L/R}_{\omega,\vec{k}}|\gamma\rangle_I=\Big|\int \extd p_z\int  \extd\bm{p}\  \alpha^{L/R}_{\omega,\vec{k};p_-,\vec{p}}\ \gamma^*_{p_-,\vec{p}}\Big|^2+\ \Big|\int \extd p_z\int  \extd\bm{p}\  \beta^{L/R}_{\omega,\vec{k};p_-,\vec{p}}\ \gamma^*_{p_-,\vec{p}}\ \Big|^2+\textrm{N}^{L/R,\textrm{vac}}_{\omega,\vec{k}},
\ee 
where $\textrm{N}^{L/R,\textrm{vac}}_{\omega,\vec{k}}$ is the expectation number (\ref{UnruhRad}). The result of the final eigenstate quantization is obtained by replacing 
$\alpha,\beta$ with $\tilde{\alpha},\tilde{\beta}$. We note that the $e^{i \frac{\vec{k}^2}{4 p_-}u_0}$ phase factor in the Bogoliubov coefficients cannot be ignored for a general superposition state regardless of the quantization scheme. In particular, that phase is $\vec{k}$ dependent and therefore cannot be absorbed into a field redefinition of the creation operators $a_{p_-,\vec{p}}$. When one considers a superposed state consisting of distinct momenta, they will therefore have an interference pattern depending on the value of $u_0$, which is therefore observable.

 \section{Comments on Hawking radiation}
\label{sec:BH}
We first review the original derivation of Hawking \cite{Hawking:1974sw} (see also the review \cite{Ford:1997hb}).
The resulting Hawking black body spectrum parallels the Unruh effect. Matter-induced supertranslations $f(\theta,\phi)$ discussed earlier in the Unruh effect are mimicked in the Hawking effect by the supertranslation of the advanced time $v \mapsto v - v_0(\theta,\phi)$ induced by the non-sphericity of matter collapse. In the following we comment on the relationship between this supertranslation and displacement memory defined at null infinity as well as the spectrum resulting from the generalization of the initial vacuum state to an excited state.

Hawking's derivation goes as follows. We consider an asymptotically flat four-dimensional black hole formed by collapse of matter. For simplicity, we only consider scalar matter. The ingoing positive frequency modes at $\mathcal I^-$ and, respectively, the outgoing positive frequency modes at $\mathcal I^+$ are denoted as
\bea
f^\text{in}_{\omega' l' m' } \sim Y_{l' m' }(\theta,\phi) e^{-i \omega' v},\qquad \qquad   f^\text{out}_{\omega l m} \sim  Y_{l m }(\theta,\phi) e^{-i \omega u},
\eea
where $v = t+ r^*$ is the advanced time at $\mathcal I^-$, $u = t- r^*$ is the retarded time at $\mathcal I^+$ and $\omega,\omega' > 0$. The scalar field is quantized  as
\bea
\phi = \int_0^{\infty} \extd\omega' \sum_{ l' m'}  \left( a^{\text{in}}_{ l' m'} f^\text{in}_{\omega' l' m'} + a^{\text{in} \dagger}_{\omega' l' m' } f^\text{in*}_{\omega' l' m' } \right) .
\eea
The in-vacuum $ | 0 \rangle_{\text{in}} $ at $\mathcal I^-$ is defined as $a^\text{in}_{\omega' l' m'} | 0 \rangle_{\text{in}} = 0$. We assume that the matter is in the vacuum state $ | 0 \rangle_{\text{in}} $ at $\mathcal I^-$. The Hawking effect arises from nontrivial Bogoliubov coefficients between the ingoing and outgoing modes
\bea
f^\text{out}_{\omega l m} = \int_0^{\infty} \extd\omega' \sum_{l'm'} \left( \alpha_{\omega l m, \omega' l' m'} f^\text{in}_{\omega' l' m'} +  \beta_{\omega l m, \omega' l' m'} f^{\text{in}*}_{\omega' l' m'}  \right).
\eea
The result is only derived for late retarded times $u$. In the semi-classical description, the late time effect is dominated by high-frequency modes at $\mathcal I^-$ that propagated through the collapsing body just before the event horizon formed. One can then use the geometrical optics approximation. Hawking gives a general ray-tracing argument that leads to the relationship\footnote{The notation $o(r)$ means subleading with respect to $r$.}
\bea
v_0 - v = e^{\kappa (u_0 - u)} + o(e^{-\kappa u} ).\label{modeconnection}
\eea
Here, $\kappa$ is the black hole surface gravity as measured by an asymptotic observer and $u_0$ is a constant. The limiting advanced time $v_0$ is the latest time that a null geodesic could leave $\mathcal I^-$, pass through the center of the body and escape to $\mathcal I^+$ without being trapped by the event horizon. For a spherically symmetric collapse, $v_0$ is a constant that is independent of the angle of emission $\theta,\phi$. This leads to a one-to-one map between the ingoing $l', m'$ and outgoing $l, m$ mode numbers. For a nonspherically symmetric collapse, the quantities $v_0=v_0(\theta,\phi)$ and $u_0=u_0(\theta,\phi)$ are determined for each null geodesic from the details of the collapse. Nonspherically symmetric collapse leads to the Bogoliubov coefficients
    \begin{subequations}
        \begin{align}
            \alpha_{\omega l m,\omega' l' m'} &=  t_{\omega l m}\times  \alpha^S_{\omega ,\omega'} \times \phi_{\omega lm, \omega' l' m'},\label{alpha}\\
            \beta_{\omega l m,\omega' l' m'} &= t_{\omega l m}\times  \beta^S_{\omega ,\omega'} \times \phi_{\omega lm, -\omega' l' -m'}(-1)^{m'},\label{beta}
        \end{align}
    \end{subequations}
where $t_{\omega l m}$ is the classical transmission coefficient for the black hole and $\alpha^S_{\omega ,\omega'}$, $\beta^S_{\omega,\omega'}$ are the spherically symmetric Bogoliubov coefficients given at large ingoing frequency $\omega'$ by  \cite{Hawking:1974sw,Hawking:1976ra}
    \begin{subequations}
        \begin{align}
            \alpha^S_{\omega ,\omega'} & \approx \frac{1}{2\pi} \sqrt{\frac{\omega'}{\omega}} \Gamma(1-\frac{i\omega}{\kappa}) (-i \omega')^{-1+\frac{i\omega}{\kappa}}, \\
            \beta^S_{\omega, \omega'} &\approx -i \alpha^S_{\omega ,-\omega'}.
        \end{align}
    \end{subequations}
The angular mixing factor is given by\footnote{Note that the derivation of Hawking \cite{Hawking:1974sw} from (2.18) to (2.19) contains a misprint. There should be no factor of $e^{i \omega v_0}$ in (2.19). See (2.19)-(2.21) of Ford \cite{Ford:1997hb} for a derivation but note that $\omega v_0$ in (2.21) should be $\omega' v_0$.}
    \begin{equation}
        \phi_{\omega lm, \omega' l' m'} = \int \extd\Omega e^{i  \omega' v_0(\theta,\phi)} Y_{lm}(\theta,\phi)Y^*_{l'm'}(\theta,\phi)e^{-i \omega u_0(\theta,\phi)},      \label{mode}
    \end{equation}
where $\extd\Omega = \frac{1}{4 \pi}\sin \theta \extd\theta \extd\phi$ is the unit measure on the sphere. The angular mixing factor can be formally deduced from the spherically symmetric case by performing the advanced supertranslation shift $v \rightarrow v - v_0(\theta,\phi) + v_0^{symm}$ and the retarded supertranslation shift $u \rightarrow u - u_0(\theta,\phi)+u_0^{symm}$ where $v_0^{symm}$, $u_0^{symm}$ are the constants arising from the spherically symmetric collapse. The mixing factor \eqref{mode} is analoguous to \eqref{eq:PhiDef} derived for Unruh radiation in the presence of an outgoing supertranslating shock wave. We would have obtained the two exponential factors in the Unruh case \eqref{eq:PhiDef} upon considering both an ingoing and an outgoing supertranslating shockwave. For both the spherically symmetric and nonspherically symmetric cases, the resulting Bogoliubov coefficients obey 
  \begin{equation}
\sum_{l'm'}|\alpha_{\omega l m,\omega' l' m'} |^2= e^{\frac{2\pi \omega}{\kappa}} \sum_{l'm'}|\beta_{\omega l m,\omega' l' m'}|^2.  \label{cond:th}
    \end{equation}
Consistency of the Bogoliubov coefficients combined with \eqref{cond:th} finally leads to the black body spectrum 
\bea\label{vacnum}
N_{\omega l m} = \int_0^\infty \extd\omega' \sum_{l',m'} |\beta_{\omega l m,\omega' l' m'}|^2 \propto \frac{1}{e^{\frac{2 \pi \omega}{\kappa}} -1}. 
\eea
The angular mixing factor $\phi_{\omega lm, \omega' l' m'}$ due to the non-sphericity of the collapse does not affect the spectrum. This is analogue to our result \eqref{UnruhRad} for the Unruh radiation with a supertranslating shockwave. The result can be straightforwardly generalized to any dimension.

Let us comment upon the relationship between the limiting advanced time $v_0(\theta,\phi)$ and displacement memory defined at null infinity. The quantity $v_0(\theta,\phi)$ is determined from the details of the non-sphericity of the collapse during the collapsing phase, and it is defined as a leading order effect in \eqref{modeconnection}. It depends upon the global properties of null rays between past and future null infinity that, in turn, depend upon the stress-tensor of matter in the bulk of spacetime. In the case of collapse caused by a shockwave sourced by a stress-tensor energy $T_{vv} = r^{-2} \rho(\theta,\phi)+o(r^{-2})$ that is determined at leading order by a given energy distribution $\rho(\theta,\phi)$ at past null infinity, one can relate the limiting advanced time $v_0(\theta,\phi)$ to the displacement memory at null infinity sourced by the energy distribution $\rho(\theta,\phi)$ thanks to Einstein's constraint equations \cite{Strominger:2014pwa,Hawking:2016msc,Compere:2016hzt}. Such displacement memory leads to a shift of the canonical variable defined at $\mathcal I^+$ that encodes the displacement memory and that is also shifted by BMS supertranslations. We conclude that the limiting advanced time $v_0(\theta,\phi)$ is a functional of the leading order matter fields and directly relates to displacement memory at null infinity, consistently with the analysis of \cite{Hawking:2016sgy,Javadinazhed:2018mle}.  In that regard, the existence of a generalization of the limiting advanced time $v_0$ in any dimension is consistent with the existence of the triangular relationship between supertranslations, displacement memory and soft theorems in any dimension at null infinity \cite{Pate:2017fgt,Campiglia:2017xkp,Mao:2017wvx}. We expect that the subleading components of the stress-tensor will encode the details of the subleading corrections in \eqref{modeconnection}. 

We noticed in Section \ref{onep} that the generalization of the Unruh effect to excited states leads to a dependence of the Rindler occupation number on the shock wave form factor. We might therefore ask whether or not the Hawking spectrum is modified if one considers an ingoing excited state instead of the ingoing vacuum $ | 0 \rangle_{\text{in}}$. Indeed, the initial state contains the very matter that collapses and the initial state is therefore not a vacuum state. This question was analyzed in \cite{Wald:1976ka}. Starting with an excited state leads to classical scattering but also to stimulated emission in addition to spontaneous Hawking emission. All modes that spontaneously emit can also be stimulated in principle. However, introducing stimulated radiation at late retarded time $u$ requires an initial energy of the order of $e^{\kappa u}$ as a direct consequence of the relationship \eqref{modeconnection} and the uncertainty principle at $\mathcal I^-$. In more detail, one requires to fine-tune the initial ingoing state to lie in a close range of advanced time $v_0 - \eps < v < v_0$ with $\eps \sim e^{-\kappa u}$ in order to reach the late time $u$. Such modes will spend large proper times around the horizon until finally leaving the horizon region and reaching $\mathcal I^+$. At $\mathcal I^-$ the uncertainty relation is $\Delta E\, e^{-\kappa u} \geq 1$ or $\Delta E \geq e^{\kappa u}$. Since initial energy is capped for physical states, there cannot be a stimulated emission at late times. The angular dependence of the limiting advanced time (or in other words the supertranslation shift) is therefore irrelevant for late time spontaneous and stimulated emission. 

Yet, scattering will in general depend upon the supertranslation shift for a nonspherical collapse. Let us take as ingoing state an eigenstate with definite mode numbers  $\omega',l',m'$. The expectation number $\textrm{N}^{\text{out}}_{\omega l m}$ at $\mathcal{I}^+$ is 
\bea
{}_{\text{in}}\langle \omega'l'm'|\textrm{N}^{\text{out}}_{\omega l m}|\omega'l'm'\rangle_{\text{in}}&=&{}_{\text{in}}\langle 0|a^{\text{in}}_{\omega'l'm'}a^{\text{out}\dagger}_{\omega l m}a^{\text{out}}_{\omega l m}a^{\text{in}\dagger}_{\omega'l'm'}|0\rangle_{\text{in}}\nn\\&=&\Big|\alpha_{\omega l m,\omega' l' m'}\Big|^2+\Big|\beta_{\omega l m,\omega' l' m'}\Big|^2+\langle\omega'l'm'|\omega'l'm'\rangle \textrm{N}_{\omega l m}^{\text{vac}},
\eea 
where $\textrm{N}^{\text{vac}}_{\omega l m}$ is the spontaneous Hawking emission (\ref{vacnum}). Similarly, for an ingoing state in a superposition
\be
|\gamma\rangle_{\text{in}}=\sum_{\omega'l'm'}\gamma_{\omega'l'm'}|\omega'l'm'\rangle_{\text{in}},\qquad \sum_{\omega'l'm'}\Big|\gamma_{\omega'l'm'}\Big|^2=1,
\ee 
the expectation number at $\mathcal{I}^+$ is 
\be
{}_{\text{in}}\langle \gamma|\textrm{N}^{\text{out}}_{\omega l m}|\gamma\rangle_{\text{in}}
=\Big|\sum_{\omega'l'm'}\alpha_{\omega l m,\omega'l'm'}\gamma^*_{\omega'l'm'}\Big|^2+\Big|\sum_{\omega'l'm'}\beta_{\omega l m,\omega'l'm'}\gamma^*_{\omega'l'm'}\Big|^2 +
\textrm{N}^{\text{vac}}_{\omega l m}.
\ee 
Since the Bogoliubov coefficients $\alpha_{\omega l m,\omega'l'm'}$ and $\beta_{\omega l m,\omega'l'm'}$ given in (\ref{alpha})-(\ref{beta}) deviate from the spherically symmetric result, we conclude that the expectation number will depend upon the limiting advanced time $v_0(\theta,\phi)$ and retarded time shift $u_0(\theta,\phi)$.  

Another relevant quantity in the scattering theory around black hole is the conditional probability $P(k | j)$ that $k$ scalar particles in a given mode $\omega,l,m$ emerge given $j$ incoming particles with  modes $\omega',l',m'$. This probability was obtained by Bekenstein and Meisel \cite{Bekenstein:1977mv} and derived from first principles by Panangaden and Wald \cite{Panangaden:1977pc} with the result
\bea
P( k | j) = \frac{(1-x)x^k (1 - |R|^2)^{j+k}}{(1-|R|^2 x)^{j+k+1}} \sum_{m=0}^{\text{min}(j,k)} \frac{(j+k-m)!}{(j-m)! (k-m)! m!} \left[ \frac{(|R|^2 - x)(1-|R|^2 x)}{(1-|R|^2)^2 x}\right]^m.
\eea
where $x = e^{-\pi \frac{\omega - m \Omega_H}{\kappa}}$, and $R$ is the reflection coefficient for the mode $\omega,l,m$. One can replace $|R|^2 = 1-|T|^2$ in terms of the transmission coefficient $T$. Here, we note that given the multiplicative factor between the classical transmission coefficient and the supertranslation shift factor $\phi_{\omega lm, \omega' l' m'}$ \eqref{mode} appearing in the Bogoliubov coefficients \eqref{alpha}, we expect that the total transmission coefficient $T$ is the product of the classical transmission coefficient times the angular mixing factor.

\section{Conclusion}\label{sec:Discussion}

We described two quantization schemes for free scalar fields in the background of a planar shock wave. The shock wave induces a supertranslation of planar waves along the null wavefront that is entirely determined by the shock wave matter stress-tensor. We considered Rindler (i.e. uniformly accelerated) observers that either cross or do not cross the shock wave. In either case and for either choice of vacuum, we obtained that the spectrum observed by the uniformly accelerated observer is the standard Unruh spectrum that is independent of the form factor characterizing the shock wave. However, we noted that excited states with respect to one global vacuum lead to Rindler occupation numbers that explicitly depend on the shock wave form factor. 

We drew a parallel between the Unruh effect in the presence of supertranslating shockwaves and the Hawking effect for nonspherical collapse. We proposed that the analogue of the matter-induced supertranslation occuring in the Unruh effect with shockwaves is the limiting advanced time shift (and the second retarded time shift) characterizing the properties of nonspherical black hole collapse. Following Hawking's derivation and Wald's subsequent treatment, we reviewed that the late spontaneous and stimulated black hole emission do not depend on the limiting advanced time and retarded time shift. However, we pointed out that the limiting advanced time and retarded time shift influences the properties of scattering (with initial excited states) because of its presence in the Bogoliubov coefficients. These properties are in complete analogy with the ones of Rindler observers in the presence of shock waves. 

Our results bring further evidence that Hawking radiation is not modified by soft hair implanted by supertranslating shock waves, at least not using the mechanisms analyzed here. It complements the results of  \cite{Javadinazhed:2018mle} obtained from the perspective of dressing states with infrared soft hair. While these results are consistent with the soft hair conjecture \cite{Strominger:2017aeh}, they do not provide a mechanism for correlating in a bijection soft and hard sectors. Even though the infrared sector can be constrained by leading and subleading conservation laws, the details of the collapse (including the relationship between the ingoing and outgoing modes at null infinity) depend upon the details of the higher subleading structure of the fields. We therefore expect that a bijection between soft and hard sectors could only be achievable by including higher subleading constraints between hard and soft sectors than previous derived. 

\vspace{-0.2cm}
\subsection*{Acknowledgements}

We thank Steve Giddings, Simeon Hellerman, Reza Javadinezhad, Uri Kol, Massimo Porrati, Philippe Spindel, Andrew Strominger and Masataka Watanabe for valuable and enlightening discussions and comments. M.R. wants to thank the Center for the Fundamental Laws of Nature and the Black Hole Initiative at Harvard University as well as the Center for Gravitational Physics and the Yukawa Institute for Theoretical Physics (YITP) at Kyoto University for their hospitality during the completion of this work. The research of G.C. and M.R. is supported by the ERC Starting Grant 335146 ``HoloBHC''. The research of M.R. is partially supported by the ERC Advanced Grant ``High-Spin-Grav" and by FNRS-Belgium (convention FRFC PDR T.1025.14 and  convention IISN 4.4503.15). The research of J.L. was supported by the Ministry of Science, ICT \&
Future Planning, Gyeongsangbuk-do and Pohang City. G.C. is FNRS Research Associate. 

\appendix
\section{Orthogonality relations}
\label{sec:ortho}

Here we prove the orthogonality conditions \eqref{rel2}-\eqref{rel3}, where we set $u_0=0$ for simplicity. The following relations will be very useful in order to evaluate the $z$-integrals appearing in the Klein-Gordon norm\footnote{Multiplications of distributions are well defined in the sense of Colombeau \cite{colombeau1990}.}
    \begin{subequations}
        \begin{align}
            \int^\infty_{-\infty} \extd z\, e^{i a z + i b \Theta(-z)} & = \pi(1+e^{ib})\delta(a) - \frac{i}{a}(e^{i b}-1),\label{conj2}\\
            \int^\infty_{-\infty} \extd z \,\delta(-z) e^{i a z + i b \Theta(-z)} & =\frac{e^{i b}-1}{ib}.\label{eq:App3}
        \end{align}
    \end{subequations}
For a Cauchy slice along $n_\mu=\partial_\mu t$ at $t=0$ the Klein-Gordon norm of the initial eigenstates $f^I_{k_-,\vec{k}}$ is given by
    \begin{align}
        (f^{*I}_{k_-,\vec{k}}, f^I_{l_-,\vec{l}})_{KG} &=  N_k N_l \int \extd z \int \extd\bm{x} \int \extd\bm{k}' \int \extd\bm{l}' e^{i(\vec{k}'+\vec{l}')\vec{x} + i(k'_z+l'_z)z} \left(l_t'-k_t' +(l_- - k_-) f(\vec{x}) \delta(-z) \right) \nonumber \\
        &\quad\times A_{k_-,\vec{k}}(\vec{k}',\Theta(-z))A_{l_-,\vec{l}}\,(\vec{l}',\Theta(-z)),
    \end{align}
where $k_z' \equiv -k_-+\frac{\vec{k}^{\prime 2}}{4k_-}$, $l_z' \equiv -l_-+\frac{\vec{l}^{\prime 2}}{4l_-}$, $k_t' \equiv k_-+\frac{\vec{k}^{\prime 2}}{4k_-}$, $l_t' \equiv l_-+\frac{\vec{l}^{\prime 2}}{4l_-}$. We split the computation in two terms: 
    \begin{subequations}
        \begin{align}
            (f^{*I}_{k_-,\vec{k}}, f^I_{l_-,\vec{l}})_{KG} &= T_1 + T_2,\\
            T_1 &= N_k N_l \int \extd z \int \extd\bm{x} \int \extd\bm{k}' \int \extd\bm{l}' e^{i(\vec{k}'+\vec{l}')\vec{x} + i(k'_z+l'_z)z} (l_t'-k_t') \nonumber \\
            &\quad\times A_{k_-,\vec{k}}(\vec{k}',\Theta(-z))A_{l_-,\vec{l}}\,(\vec{l}',\Theta(-z)),\\
            T_2 &=  N_k N_l \int \extd z \int \extd\bm{x} \int \extd\bm{k}' \int\extd\bm{l}' e^{i(\vec{k}'+\vec{l}')\vec{x} + i(k'_z+l'_z)z} (l_- - k_-) f(\vec{x}) \delta(-z)  \nonumber \\
            &\quad\times A_{k_-,\vec{k}}(\vec{k}',\Theta(-z))A_{l_-,\vec{l}}\,(\vec{l}',\Theta(-z)).
        \end{align}
    \end{subequations}
For $T_1$ we can perform the $\int \extd\bm{x}$ integral first yielding a delta distribution $(2\pi)^{D-2}\delta(\vec{k}'+\vec{l}')$ and after evaluating the $\int \extd\bm{l}'$ integral we can set $\vec{l}'=-\vec{k}$. We then express $A_{k_-,k}$ and  $A_{l_-,l}$ using their definition \eqref{eq:PhiDef}, and perform the integral $\int \extd z$. Using \eqref{conj2} one obtains again two terms
    \begin{equation}
        T_1 =  T_1'+T_1'',
    \end{equation}
with
    \begin{subequations}
        \begin{align}
            T_1' &=  \frac{N_k N_l}{(2\pi)^{D-2}} \int \extd\bm{x}' \int \extd\bm{x}''  \int \extd\bm{k}' e^{i(\vec{k}-\vec{k}')\vec{x}' +i(\vec{l}+\vec{k}')\vec{x}'' } (l_t'-k_t')\pi\left(1+e^{ib}\right)\delta(a),\\
            T_1'' &= - \frac{N_k N_l}{(2\pi)^{D-2}}\int \extd\bm{x}' \int \extd\bm{x}''  \int \extd\bm{k}' e^{i(\vec{k}-\vec{k}')\vec{x}' +i(\vec{l}+\vec{k}')\vec{x}'' }(l_t'-k_t') \frac{i}{a}\left(e^{i b}-1\right),
        \end{align}
    \end{subequations}
where $a=k_z'+l_z'$ and $b=k_- f(\vec{x}')+l_- f(\vec{x}'')$. Evaluating $\int \extd\bm{k}'$ in $T_1'$ one finds that this term is generically zero since $k_z'+l_z'=0$ (after setting $\vec{l}'=-\vec{k}$) implies that $(l_t'-k_t')=0$. $T_1''$ can be further simplified by noticing that (again, after setting $\vec{l}'=-\vec{k}$)
    \begin{equation}
        \frac{l_t'-k_t'}{k_z'+l_z'}=-\frac{l_--k_-}{l_-+k_-},
    \end{equation}
and in addition performing $\int \extd\bm{k}'\int \extd\bm{x}''$. This yields
    \begin{equation}
        T_1 =i\left(\frac{l_--k_-}{l_-+k_-}\right)N_k N_l\int \extd\bm{x}'e^{i(\vec{k}+\vec{l})\vec{x}'}\left(e^{if(\vec{x}')(l_-+k_-)}-1\right).
    \end{equation}
In order to evaluate $T_2$ we first evaluate $\int \extd z$ using \eqref{eq:App3} to obtain
    \begin{align}
        T_2 &=  -i(l_- - k_-) \frac{N_k N_l}{(2\pi)^{2(D-2)}} \int \extd\bm{k}' \int \extd\bm{l}' \int \extd\bm{x}\int \extd\bm{x}' \int \extd\bm{x}'' e^{i(\vec{k}-\vec{k}')\vec{x}'+i(\vec{l}-\vec{l}')\vec{x}''+i(\vec{k}'+\vec{l}')\vec{x}}f(\vec{x}) \nonumber\\
        &\quad \times  \frac{e^{ib}-1}{b}.
    \end{align}
One can then perform $\int \extd\bm{k}' \int \extd\bm{l}'$ yielding $(2\pi)^{2(D-2)}\delta(\vec{x}-\vec{x}')\delta(\vec{x}-\vec{x}'')$ and after evaluating $\int \extd\bm{x}' \int \extd\bm{x}''$ this expression simplifies to
    \begin{equation}
        T_2 = -i\left(\frac{l_--k_-}{l_-+k_-}\right)N_k N_l\int \extd\bm{x}\,e^{i(\vec{k}+\vec{l})\vec{x}}\left(e^{if(\vec{x})(l_-+k_-)}-1\right),
    \end{equation}
which immediately shows that $T_1+T_2=0$ and thus also confirming $(f^{*I}_{k_-,\vec{k}}, f^I_{l_-,\vec{l}})_{KG} = 0$.

The next step is to compute $(f^{L}_{k_-,\vec{k}}, f^I_{l_-,\vec{l}})_{KG}$. Similar to the calculation done before one finds
    \begin{align}
        (f^{L}_{k_-,\vec{k}}, f^I_{l_-,\vec{l}})_{KG} &=  N_k N_l \int \extd z \int \extd\bm{x} \int \extd\bm{k}' \int \extd\bm{l}' e^{i(\vec{l}'-\vec{k}')\vec{x} + i(l'_z-k'_z)z} \left(l_t'+k_t' +(l_- + k_-) f(\vec{x}) \delta(-z) \right) \nonumber \\
        &\quad\times A^*_{k_-,\vec{k}}(\vec{k}',\Theta(-z))A_{l_-,\vec{l}}\,(\vec{l}',\Theta(-z)).
    \end{align}
This expression can again be split into two terms
    \begin{subequations}
        \begin{align}
            (f^{L}_{k_-,\vec{k}}, f^I_{l_-,\vec{l}})_{KG} &= T_1 + T_2,\\
            T_1 &=  N_k N_l \int \extd z \int \extd\bm{x} \int \extd\bm{k}' \int \extd\bm{l}' e^{i(\vec{l}'-\vec{k}')\vec{x} + i(l'_z-k'_z)z} (l_t'+k_t') \nonumber \\
            & \quad\times A^*_{k_-,\vec{k}}(\vec{k}',\Theta(-z))A_{l_-,\vec{l}}\,(\vec{l}',\Theta(-z)),\\
            T_2 &=   N_k N_l \int \extd z \int \extd\bm{x} \int \extd\bm{k}' \int \extd\bm{l}' e^{i(\vec{l}'-\vec{k}')\vec{x} + i(l'_z-k'_z)z} (l_- + k_-) f(\vec{x}) \delta(-z)\nonumber \\
            &\quad\times A^*_{k_-,\vec{k}}(\vec{k}',\Theta(-z))A_{l_-,\vec{l}}\,(\vec{l}',\Theta(-z)).
        \end{align}
    \end{subequations}
For $T_1$ we can perform $\int \extd\bm{x} \int \extd\bm{l}'$ which sets $\vec{l}'=\vec{k}$. We then express $A_{k_-,k}$ and  $A_{l_-,l}$ using their definition \eqref{eq:PhiDef}, and perform $\int \extd z$. With the help of \eqref{conj2} one obtains again two terms
    \begin{equation}
        T_1 =  T_1'+T_1'',
    \end{equation}
with
    \begin{subequations}
        \begin{align}
            T_1' &=  \frac{N_k N_l}{(2\pi)^{D-2}} \int \extd\bm{x}' \int \extd\bm{x}''  \int \extd\bm{k}' e^{i(\vec{k}-\vec{k}')\vec{x}' +i(\vec{l}-\vec{k}')\vec{x}'' } (l_t'+k_t')\pi\left(1+e^{i\tilde{b}}\right)\delta(\tilde{a}),\\
            T_1'' &= - \frac{N_k N_l}{(2\pi)^{D-2}}\int \extd\bm{x}' \int \extd\bm{x}''  \int \extd\bm{k}' e^{i(\vec{k}-\vec{k}')\vec{x}' +i(\vec{l}-\vec{k}')\vec{x}'' }(l_t'+k_t') \frac{i}{\tilde{a}}\left(e^{i \tilde{b}}-1\right),
        \end{align}
    \end{subequations}
where $\tilde{a}=l_z'-k_z'$ and $\tilde{b}=l_- f(\vec{x}'')-k_- f(\vec{x}')$. In contrast to the case before $T_1'$ is not always zero. This can be seen by looking at $\tilde{a}=(k_--l_-)(1+\frac{\vec{k}^{\prime 2}}{4k_-l_-})$. For $\vec{k}^{\prime 2}\in\mathbb{R}$ this expression is never zero. Only for $k_-=l_-$ this expression vanishes. Thus we can rewrite $\delta(\tilde{a})$ as $\frac{\delta(l_--k_-)}{1+\frac{\vec{k}^{\prime 2}}{4k_-l_-}}$. This gives
    \begin{equation}
        T_1' =\frac{N_k N_l}{(2\pi)^{D-2}} \int \extd\bm{x}' \int \extd\bm{x}''  \int \extd\bm{k}' e^{i(\vec{k}-\vec{k}')\vec{x}' +i(\vec{l}-\vec{k}')\vec{x}'' } (l_-+k_-)\pi\left(1+e^{i\tilde{b}}\right)\delta(l_--k_-).
    \end{equation}
Since this expression is only non-zero for $l_-=k_-$ it can be further simplified to
    \begin{equation}
        T_1' = k_- \frac{N_k^2}{(2\pi)^{D-3}} \int \extd\bm{x}' \int \extd\bm{x}''  \int \extd\bm{k}' e^{i(\vec{k}-\vec{k}')\vec{x}' +i(\vec{l}-\vec{k}')\vec{x}'' }\left(1+e^{ik_-\left(f(\vec{x}'')-f(\vec{x}')\right)}\right)\delta(l_--k_-).
    \end{equation}
Evaluating first $\int \extd\bm{k}'$ yields $(2\pi)^{D-2}\delta(\vec{x}'-\vec{x}'')$ and then $\int \extd\bm{x}''$ this integral simplifies to
    \begin{equation}
        T_1' = 2k_- N_k^2(2\pi)^{D-1} \int \extd\bm{x}' e^{i(\vec{l}-\vec{k}')\vec{x}' }\delta(l_--k_-).
    \end{equation}
After evaluating the last integral and using that
    \begin{equation}
        N_k=\frac{1}{\sqrt{(2\pi)^{D-1} 2k_t}},
    \end{equation}
one obtains for $T_1'$
    \begin{equation}
        T_1'=\delta(\vec{l}-\vec{k})\delta(l_z-k_z).
    \end{equation}
Here we used the relations 
    \begin{align}
        k_t&=k_-+k_+\nn,\\
        k_z&=k_+-k_-,\\
        \delta(l_z-k_z)\delta(\vec{l}-\vec{k})&=\frac{k_-}{k_++k_-}\delta(l_--k_-)\delta(\vec{l}-\vec{k}).\nn
    \end{align}
$T_1''$ can again be determined by first evaluating $\int \extd\bm{k}'\int \extd\bm{x}''$. This yields
    \begin{equation}
        T_1'' =i\left(\frac{l_-+k_-}{l_--k_-}\right)N_k N_l\int \extd\bm{x}'e^{i(\vec{l}-\vec{k})\vec{x}'}\left(e^{if(\vec{x}')(l_--k_-)}-1\right).
    \end{equation}
In order to compute $T_2$ we first evaluate $\int \extd z$ using \eqref{eq:App3} to obtain
    \begin{align}
        T_2 &=  -i(l_- + k_-) \frac{N_k N_l}{(2\pi)^{2(D-2)}} \int \extd\bm{k}' \int \extd\bm{l}' \int \extd\bm{x}\int \extd\bm{x}' \int \extd\bm{x}'' e^{i(\vec{k}-\vec{k}')\vec{x}'+i(\vec{l}-\vec{l}')\vec{x}''+i(\vec{l}'-\vec{k}')\vec{x}}f(\vec{x}) \nonumber\\
        &\quad \times  \frac{e^{i\tilde{b}}-1}{\tilde{b}}.
    \end{align}
One can then perform $\int \extd\bm{k}' \int \extd\bm{l}'$ yielding $(2\pi)^{2(D-2)}\delta(\vec{x}-\vec{x}')\delta(\vec{x}-\vec{x}'')$ and after evaluating $\int \extd\bm{x}' \int \extd\bm{x}''$ this expression simplifies to
    \begin{equation}
        T_2 = -i\left(\frac{l_-+k_-}{l_--k_-}\right)N_k N_l\int \extd\bm{x}\,e^{i(\vec{l}-\vec{k})\vec{x}}\left(e^{if(\vec{x})(l_--k_-)}-1\right).
    \end{equation}
Putting together $T_1+T_2$ one finds that
    \begin{equation}
        (f^{I}_{k_-,\vec{k}}, f^I_{l_-,\vec{l}})_{KG}=\delta(\vec{l}-\vec{k})\delta(l_z-k_z).
    \end{equation}
The result for $(f^{*I}_{k_-,\vec{k}}, f^{*I}_{l_-,\vec{l}})_{KG}$ can be straightforwardly obtained in a similar manner to yield
    \begin{equation}
        (f^{*I}_{k_-,\vec{k}}, f^{*I}_{l_-,\vec{l}})_{KG}=-(f^{I}_{k_-,\vec{k}}, f^I_{l_-,\vec{l}})_{KG}=-\delta(\vec{l}-\vec{k})\delta(l_z-k_z).
    \end{equation}


\providecommand{\href}[2]{#2}\begingroup\raggedright\endgroup

\end{document}